\documentclass[preprint2]{aastex62}

\newcommand\Gaia{\emph{Gaia}}
\newcommand{\kmsec}{km~s$^{-1}$}

\newcommand\mumod{\bm{\mu_{\rm mod}}}

\newcommand\cOne{C1-L1551}
\newcommand\cTwo{C2-L1495}
\newcommand\cThreeOG{C3-L1517}
\newcommand\cThree{C3-L1517-Halo}
\newcommand\cFour{C4-L1517-Center}
\newcommand\cFive{C5-L1546}
\newcommand\cSix{C6-L1524}
\newcommand\cSevenOG{C7-L1527-B213}
\newcommand\cSeven{C7-L1527}
\newcommand\cEight{C8-B213}
\newcommand\cNineOG{C9-118Tau}
\newcommand\cNine{C9-118TauE}
\newcommand\cTen{C10-118TauW}

\newcommand\dOne{D1-L1544}
\newcommand\dTwo{D2-L1558}
\newcommand\dThree{D3-South}
\newcommand\dFour{D4-North}
\newcommand\dFive{D5-OutlierCentral}
\newcommand\dSix{D6-OutlierNorth}
\newcommand\dSeven{D7-OutlierEast}

\usepackage{bm}
\usepackage{amsmath}

\graphicspath{{./}{figures/}}

\received{5 May 2021}
\revised{20 May 2021}
\accepted{24 May 2021}
\submitjournal{AJ}

\shorttitle{Substructure of Taurus in Gaia EDR3}
\shortauthors{Krolikowski et al.}

\begin{document}

\title{\Gaia\ EDR3 Reveals the Substructure and Complicated Star Formation History of the Greater Taurus-Auriga Star Forming Complex}

\correspondingauthor{Daniel M. Krolikowski}
\email{krolikowski@utexas.edu}

\author[0000-0001-9626-0613]{Daniel M. Krolikowski}
\author[0000-0001-9811-568X]{Adam L. Kraus}
\author[0000-0001-9982-1332]{Aaron C. Rizzuto}
\affil{Department of Astronomy, University of Texas at Austin, 2515 Speedway, Stop C1400, Austin, TX 78751, USA}

\begin{abstract}
The Taurus-Auriga complex is the prototypical low-mass star forming region, and provides a unique testbed of the star formation process, which left observable imprints on the spatial, kinematic, and temporal structure of its stellar population. Taurus's rich observational history has uncovered peculiarities that suggest a complicated star forming event, such as members at large distances from the molecular clouds and evidence of an age spread. With \Gaia, an in-depth study of the Taurus census is possible to confirm membership, identify substructure, and reconstruct its star formation history. We have compiled an expansive census of the greater Taurus region, identifying spatial subgroups and confirming that Taurus is substructured across stellar density. There are two populations of subgroups: clustered groups near the clouds and sparse groups spread throughout the region. The sparse groups comprise Taurus's distributed population, which is on average older than the population near the clouds, and hosts sub-populations up to 15~Myr old. The ages of the clustered groups increase with distance, suggesting that the current star formation was triggered from behind. Still, the region is kinematically coherent, and its velocity structure reflects an initial turbulent spectrum similar to Larson's Law that has been modified by dynamical relaxation. Overall, Taurus has a complicated star formation history, with at least two epochs of star formation featuring both clustered and distributed modes. Given the correlations between age and spatial distribution, Taurus might be part of a galaxy-scale star forming event that can only begin to be understood in the \Gaia\ era.
\end{abstract}


\section{Introduction}\label{sec:introduction}

Star formation is a galaxy-scale event. The cold interstellar medium can become dense enough to form turbulent molecular clouds, which are balanced against global collapse from self-gravity by support from internal pressures, such as from turbulence \citep{Bonazzola87,Hennebelle12} and magnetic fields \citep{Li06,Crutcher10}. Turbulence, however, can also create local overdensities that become dense enough for gravity to overtake the supporting pressures, leading to fragmentation and collapse \citep{Hopkins13} into cores that eventually form stars. Star-forming structures are now observed across this full range of scales, including hundreds of parsecs long filaments \citep{Goodman14,Zucker19} to a 3~kpc gas wave encompassing most nearby star forming regions \citep[the Radcliffe wave;][]{Alves20}, all of which contain structure down to the core-scale.

The star formation process leaves lasting observable imprints on nascent stellar populations. The spatial, kinematic, and temporal structure of these stellar populations illustrates the initial conditions in their parent molecular clouds, and directly constrains their star formation history. Understanding the conditions in molecular clouds is crucial in determining the relative importance and role of dominant physical processes in star formation. Knowledge of their star formation history is needed to model the evolution of stellar populations en masse and place them into the greater galactic star formation context.

There are luminosity spreads observed in many young clusters, which through an HR diagram interpretation imply age spreads \citep{Pecaut12,Soderblom14,Rizzuto16}. These measurements are complicated, however, by extinction, disks, accretion, and unresolved binarity, all of which can cause a luminosity spread without any finite spread in age \citep{Jeffries11,Sullivan21}. If these age spreads are real, however, they imply ubiquitous complicated star formation histories. Age spreads have been explained by extended star formation timescales in these regions, significantly longer than the free-fall time, with many regions experiencing accelerated star formation rates over time \citep{Palla00,Caldwell18}. In a specific example, the age spreads in Upper Sco can be explained with an extended period of constant star formation followed by a short burst of intense star formation both started and ended by supernovae \citep{Fang17}.

Molecular clouds are highly substructured, featuring regions of higher density gas in various spatial configurations, particularly along filaments \citep{Andre10,Molinari10}. It is thought that stars may form as gas is funneled along these filaments to reach the critical density for fragmentation, and if so, nascent stellar populations should trace the parent filament or at the very least be clustered \citep{Lada03,Enoch07,Roccatagliata20}. It is also found that stars do not exclusively form in high density clusters, and may form in a distributed mode at low stellar density, although this does not preclude substructure \citep{Bressert10}. It is important to tease out the relative importance of various modes of star formation, and in particular how and if they coexist in the same region. The spatial structure and density of stellar populations at young ages helps to set the timescale for spatial dissolution, and how it may be affected by the initial kinematic structure which may persist for hundreds of millions of years \citep{Mamajek16}. 

Turbulence sets the velocity field of the molecular cloud, which results in a power-law relation with higher velocity dispersion at larger spatial scales \citep{Larson81,Solomon87}. The slope and amplitude of this relation is directly related to the turbulent energy present in the cloud and the relation between the turbulence and the gas (e.g. subsonic or supersonic, incompressible or compressible) \citep{McKee07}. This velocity field should be inherited by newly formed stars and reflected in their velocity structure function at birth, before dynamical evolution can occur. Recent work from \citet{Ha21} has studied the velocity structure of the stellar population of Orion, successfully using it to draw conclusions about the turbulent nature of the molecular clouds and to infer dynamical evolution of the stellar population between birth and the present day.

One young stellar population whose properties can be mapped to study the star formation process is the greater Taurus-Auriga complex, hereafter called Taurus. Taurus is the prototypical low-mass star forming region, the closest region of ongoing star formation in the Milky Way, and has been studied in detail for nearly a century \citep{Burnham1890,Barnard27,Joy45}. The Taurus clouds have substructure, featuring multiple filaments of molecular gas \citep{Goldsmith08} which are well populated with forming or newly-formed stars \citep{Kenyon95}. Disk-bearing Taurus members are largely concentrated near the sites of ongoing star formation, and their census is likely near completion \citep{Luhman10,Rebull10,Rebull11,Esplin14}. There are, however, disk bearing objects that sit isolated from the clouds, such as T Tauri itself, indicating a more complicated star forming history.

There is also a population of young, disk-free objects distributed throughout the Taurus region at large angular separations from the clouds whose census is almost certainly incomplete \citep{Wichmann96,Wichmann00,Slesnick06b,GomezDeCastro15}. It is suggested that these objects cannot have formed directly with the assuredly young, disk-bearing clustered objects, indicating either a previous generation of star formation or a distributed mode of star formation in the region. A comprehensive re-analysis of all disk-free candidate members by \citet{Kraus17} suggests that many of these stars are indeed young and broadly comoving with the youngest Taurus members, but also that this population is older ($\tau \ga 10$~Myr) and closer ($d \la 120$~pc). \citet{Zhang18} has also found a substantial new population of low-mass objects that are members of this population, demonstrating they outline a full IMF. The occurrence of multiple generations of star formation in Taurus would affect ongoing debate regarding molecular cloud lifetimes \citep{Krumholz07,Murray11,Federrath15} and require a more complicated view of the Taurus cloud complex.

There has been recent work using \Gaia\ DR2 to refine the boundaries of Taurus and identify substructure within the region. \citet{Luhman18} refined the census of Taurus using the \Gaia\ DR2 data, rectified the previously abnormal initial mass function found in the region, and claimed that the older population proposed by \citet{Kraus17} is distinct from the canonical Taurus region. \citet{Galli19} and \citet{Roccatagliata20} use different methods to identify substructure in the canonical Taurus region with the \Gaia\ 5D astrometric positions and velocities. They find that the region is highly structured on sub-parsec to tens of parsecs scales with bulk kinematic coherence, and that the region is not expanding.

As Taurus is a testbed for the theories of low-mass star formation, it is crucial to understand its various complexities, which needs as full of a census as possible. In the era of \Gaia, we have high precision, full astrometry for a vast majority of Taurus members -- both known and yet to be identified. With precision astrometry of the known Taurus population, we can confirm membership, identify spatial and kinematic substructure in the region, and when combined with ages, construct the tapestry that is the region's star forming history. 

In this paper, we have compiled the most complete and inclusive census of the greater Taurus region to date, and have cross-matched with the \Gaia\ EDR3 catalog to identify spatial substructure in the region. In Section~\ref{sec:census} we describe the construction of our census and compilation of various literature data. In Section~\ref{sec:substructure} we perform a clustering analysis on our sample to identify spatial substructure in the region and assign membership for our census to the identified subgroups. We describe the larger spatial distribution and structure of the Taurus region, and discuss the spatial characteristics of the subgroups in Section~\ref{sec:spatial_results}. In Section~\ref{sec:kinematics_results} we discuss the kinematics of the region, looking at the bulk motions of the subgroups and the velocity spectrum of the full census. In Section~\ref{sec:ages} we derive ages for the subgroups and discuss implications for the star formation history of Taurus. Finally, we present conclusions about the star formation process in Taurus and summarize our results in Section~\ref{sec:discussion}.

\section{An updated Taurus census using Gaia early data release 3}\label{sec:census}

\subsection{Sample construction and Gaia crossmatch}\label{subsec:sample}

To construct a large, inclusive sample of Taurus, we have compiled a set of 658 objects from multiple sources: an internal list of canonically accepted disk-bearing Taurus members \citep[][among others]{Kenyon95,Luhman10,Rebull10,Kraus11,Kraus12,Esplin14}, a compilation of candidate class III members from \citet[][objects with a membership assessment of ``Y" or ``Y?"]{Kraus17}, and recent work on the Taurus census from \citet{Luhman18}, \citet{Esplin19}, and \citet{Galli19}. We do not include the new objects from \citet{Zhang18} as we finalized the sample based on which objects had sources in \Gaia\ DR2. We separate the HBC 358 system included in \citet{Galli19} into two objects (HBC 358 A and B), as they are widely separated enough to have separate \Gaia\ catalog entries\footnote{HBC 358 C appears to be a background object from its \Gaia\ EDR3 catalogue entry}. We also compile information on binary companions for the canonically accepted Taurus members, adding companions wide enough to have their own \Gaia\ catalog entries, but excluding any companions that are not resolved in \Gaia\ from the presented census. This includes 63 systems that contain unresolved multiples. Binary status, system architecture, and literature references are included in Table~\ref{tab:census} for objects where applicable.


For each object in our sample, we queried the \Gaia\ EDR3 catalogue \citep{GaiaEDR3} for sources within 5\arcsec\ of an object's position. For searches resulting in multiple matches, we used the astrometry and photometry of the sources found to match the object correctly. This process was straightforward for single stars, as nearly all are bright enough to have \Gaia\ measured astrometry, and incorrectly matched sources had wildly discrepant astrometry or photometry. This census cross match was first performed with the \Gaia\ DR2 catalogue \citep{GaiaDR2}, and we used the \Gaia\ DR2 to EDR3 designation matching catalogue as another test to make sure sources were matched correctly.

The cross match for \Gaia\ resolved binary systems was more complicated, as the closest match in the \Gaia\ catalogue is not necessarily the correct object match. We compared the separation and position angle between the \Gaia\ sources to literature values to ensure it is the correct pair of objects. We then assigned a match based on the position angle and \Gaia\ photometry, comparing to literature contrast as an extra check.

\newpage

\startlongtable
\begin{deluxetable}{ll}
\tabletypesize{\scriptsize}
\tablecaption{\Gaia\ EDR3 Census of the Taurus-Auriga Complex\label{tab:census}}
\tablehead{ \colhead{Column Label} & \colhead{Description } }
\startdata
Gaia & Gaia EDR3 source ID\\
Name & Common or other name\\
2MASSJ & 2MASS Point Source Catalog ID\\
R.A. & \Gaia\ Right ascension J2016 (deg)\\
Decl. & \Gaia\ Declination J2016 (deg)\\
Gmag & Gaia EDR3 $G$ magnitude (mag)\\
Internal & In internal list\\
Kraus17 & In \citet{Kraus17}\\
Luhman18 & In \citet{Luhman18}\\
Esplin19 & In \citet{Esplin19}\\
Galli19 & In \citet{Galli19}\\
Jmag & $J$ magnitude (mag)\\
Jmag Ref\tablenotemark{a} & $J$ magnitude reference\\
Jmag Flag & Flag on $J$ magnitude\\
RV & Radial velocity (\kmsec)\\
e\_RV & Radial velocity uncertainty (\kmsec)\\
RV Ref\tablenotemark{b} & Radial velocity reference\\
SpT & Spectral type\\
e\_SpT & Spectral type uncertainty\\
SpT Ref\tablenotemark{c} & Spectral type reference\\
SpT Flag & Flag on spectral type\\
A\_mag & Extinction (mag)\\
e\_A\_mag & Extinction uncertainty (mag)\\
A\_mag band & Extinction band\\
A\_mag Ref\tablenotemark{d} & Extinction reference\\
$X$ & Galactic $X$ coordinate (pc)\\
$Y$ & Galactic $Y$ coordinate (pc)\\
$Z$ & Galactic $Z$ coordinate (pc)\\
$U$ & Galactic $U$ velocity (\kmsec)\\
$V$ & Galactic $V$ velocity (\kmsec)\\
$W$ & Galactic $W$ velocity (\kmsec)\\
Group Label & Assigned group from GMM\\
Binary & Binary flag\\
Sys arc & Description of system architecture\\
Binary ref\tablenotemark{e} & Binary system reference\\
\enddata
\tablenotetext{a}{\scriptsize $J$ mag references: (1) \citet{2MASS_catalog}, (2) \citet{Esplin19}}
\tablenotetext{b}{\scriptsize RV references: (1) \citet{Wilson53}, (2) \citet{Hartmann86}, (3) \citet{Walter88}, (4) \citet{Reipurth90}, (5) \citet{Mathieu97}, (6) \citet{Sartoretti98}, (7) \citet{Reid99}, (8) \citet{Wichmann00}, (9) \citet{Muzerolle03}, (10) \citet{White03}, (11) \citet{Gontcharov06}, (12) \citet{White07}, (13) \citet{Scelsi08}, (14) \citet{Nguyen12}, (15) \citet{Torres13}, (16) \citet{Guo15}, (17) \citet{Luo15}, (18) \citet{Zhong15}, (19) \citet{Kraus17}, (20) \citet{Gagne18}, (21) \citet{GaiaDR2}, (22) \citet{Kiman19}, (23) \citet{Kounkel19}, (24) \citet{Luo19}, (25) \citet{Zhong19}, (26) This work.}
\tablenotetext{c}{\scriptsize Spectral type references: (1) \citet{Carr87}, (2) \citet{Walter88}, (3) \citet{Kenyon95}, (4) \citet{Wichmann96}, (5) \citet{Briceno99}, (6) \citet{Li98}, (7) \citet{Luhman98}, (8) \citet{Briceno99}, (9) \citet{Duchene99}, (10) \citet{Gizis99}, (11) \citet{White99}, (12) \citet{Luhman00}, (13) \citet{Martin00}, (14) \citet{Martin01}, (15) \citet{Briceno02}, (16) \citet{Duchene02}, (17) \citet{Prato02}, (18) \citet{Hartigan03}, (19) \citet{Luhman03}, (20) \citet{Walter03}, (21) \citet{White03}, (22) \citet{Abt04}, (23) \citet{Luhman04}, (24) \citet{White04}, (25) \citet{Guieu06}, (26) \citet{Itoh05}, (27) \citet{White05}, (28) \citet{Luhman06a}, (29) \citet{Luhman06b}, (30) \citet{Slesnick06b}, (31) \citet{Beck07}, (32) \citet{Scelsi08}, (33) \citet{Kraus09}, (34) \citet{Luhman09b}, (35) \citet{Prato09}, (36) \citet{Schaefer09}, (37) \citet{Connelley10}, (38) \citet{Luhman10}, (39) \citet{Rebull10}, (40) \citet{Dahm11}, (41) \citet{Nguyen12}, (42) \citet{Andrews13}, (43) \citet{Mooley13}, (44) \citet{Aberasturi14}, (45) \citet{Davies14}, (46) \citet{Esplin14}, (47) \citet{Herczeg14}, (48) \citet{Bowler15}, (49) \citet{Esplin17}, (50) \citet{Kraus17}, (51) \citet{Luhman17}, (52) \citet{Luhman18}, (53) \citet{Esplin19}}
\tablenotetext{d}{\scriptsize Extinction references: (1) \citet{Carr87}, (2) \citet{Kenyon95}, (3) \citet{Briceno98}, (4) \citet{Luhman98}, (5) \citet{Briceno99}, (6) \citet{White99}, (7) \citet{Martin00}, (8) \citet{Briceno02}, (9) \citet{Hartigan03}, (10) \citet{White03}, (11) \citet{White04}, (12) \citet{Itoh05}, (13) \citet{Guieu06}, (14) \citet{Beck07}, (15) \citet{Scelsi08}, (16) \citet{Kraus09}, (17) \citet{Luhman09b}, (18) \citet{Prato09}, (19) \citet{Connelley10}, (20) \citet{Rebull10}, (21) \citet{Esplin14}, (22) \citet{Herczeg14}, (23) \citet{Kraus17}, (24) \citet{Luhman17}, (25) \citet{Bai18}, (26) \citet{Anders19}, (27) \citet{Esplin19}}
\tablenotetext{e}{\scriptsize Binary references: (1) \citet{Leinert89}, (2) \citet{Mathieu89}, (3) \citet{Chen90}, (4) \citet{Leinert93}, (5) \citet{Ghez93}, (6) \citet{Welty95}, (7) \citet{Simon95}, (8) \citet{Simon96}, (9) \citet{Ghez97}, (10) \citet{Briceno98}, (11) \citet{Sartoretti98}, (12) \citet{Woitas98}, (13) \citet{Padgett99}, (14) \citet{Duchene99}, (15) \citet{Richichi99}, (16) \citet{Mathieu00}, (17) \citet{White01}, (18) \citet{Mason01}, (19) \citet{Duchene02}, (20) \citet{Smith05}, (21) \citet{Itoh05}, (22) \citet{White05}, (23) \citet{Kraus06}, (24) \citet{Correia06}, (25) \citet{Boden07}, (26) \citet{Duchene07}, (27) \citet{Konopacky07}, (28) \citet{Kraus07}, (29) \citet{Ireland08}, (30) \citet{Luhman09a}, (31) \citet{Kraus11}, (32) \citet{Dahm11}, (33) \citet{Kraus12}, (34) \citet{Daemgen15}}
\end{deluxetable}

\begin{figure*}
\begin{center}
\includegraphics[width=\textwidth]{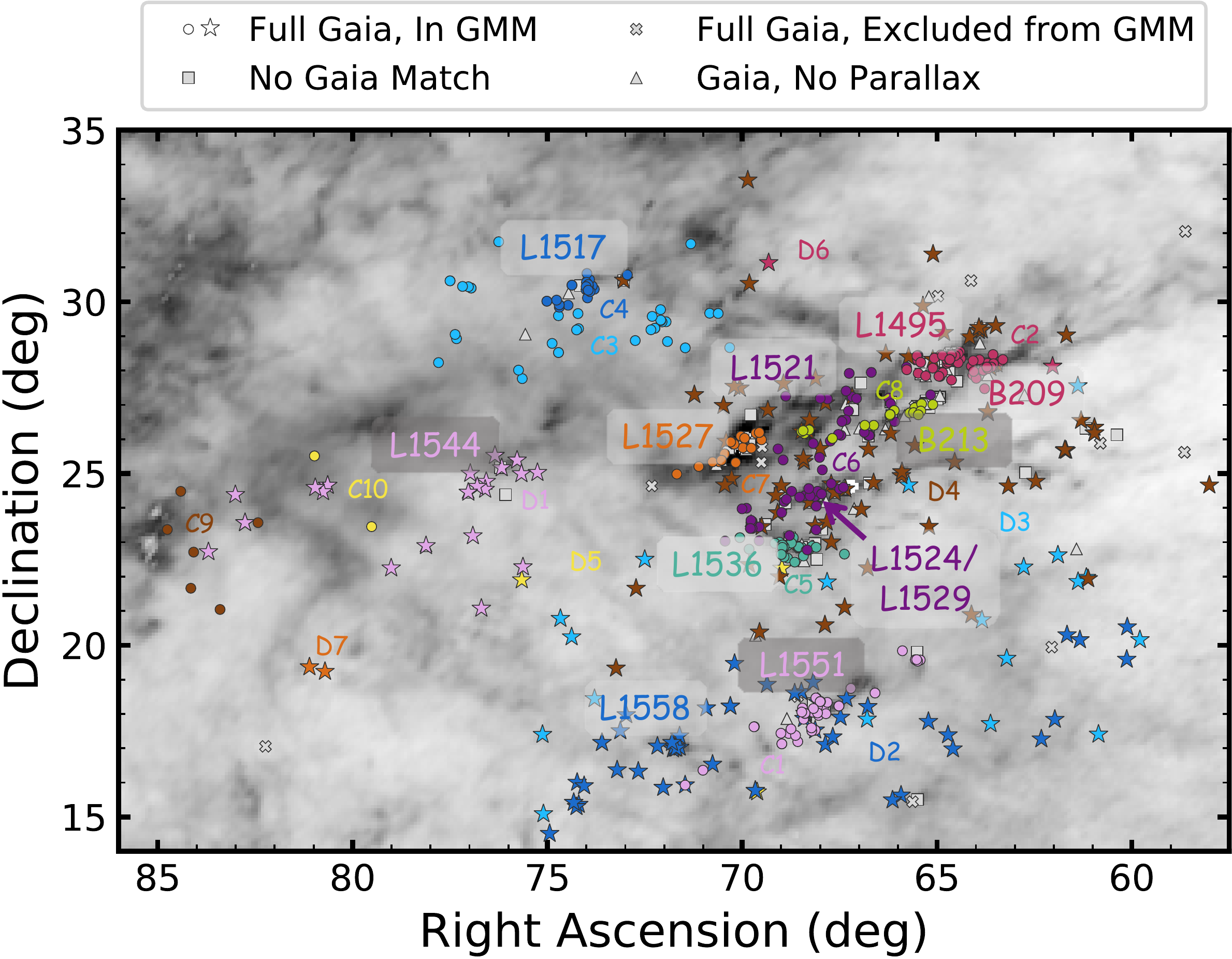}
\caption{On-sky distribution of our census of the greater Taurus-Auriga region. Objects with full \Gaia\ astrometric solutions that are included in the clustering analysis are plotted as circles and stars (to distinguish core and distributed groups defined in Section~\ref{subsec:gmm_results}). Objects with full \Gaia\ astrometric solutions that are excluded from the clustering analysis are plotted as gray X's, objects with \Gaia\ counterparts but not full astrometric solutions are plotted as gray triangles, and objects without \Gaia\ counterparts are plotted as gray squares. The colors and markers of the objects included in the clustering analysis denote their subgroup membership, which is described in Section~\ref{subsec:gmm_results} and are shown in the legend of Figure~\ref{fig:xyz_min_gmm}. The dust reddening map from \citet{Schlafly14} is plotted in the background. Important dark clouds from \citet{Barnard27} and \citet[][as compiled by \citealt{Kenyon08}]{Lynds62} are marked near their location, with their text colored the same as the group which is most coincident with the cloud. Some clouds, such as L1544, L1558, and B213, contain members of groups that extend well beyond the cloud, while other clouds, such as L1521, enclose members of multiple groups. Some groups, like Groups \cTwo\ and \cSix, comprise multiple clouds. \label{fig:sky_plot}}
\end{center}
\end{figure*}

In all, 587 objects in our census have \Gaia\ EDR3 source detections (compared to 568 for DR2), with 528 objects having full astrometric solutions (compared to 508 for DR2). The objects with \Gaia\ EDR3 counterparts are listed in Table~\ref{tab:census}. The 71 objects included in the initial census that do not have \Gaia\ EDR3 sources are listed in Table~\ref{tab:nogaia}. Figure~\ref{fig:sky_plot} shows the sky positions of all objects in our census of the greater Taurus region. Of the 104 known binary pairs in our census that have literature separation and contrast values (including pairs within higher order systems), 37 have source detections for both components and 26 have full astrometric solutions for both components. All but one system with separations above 1\arcsec\ are resolved in \Gaia\ EDR3, which we adopt as the resolution limit for Gaia EDR3. Systems with separations smaller than 1\arcsec\ that are resolved are all close to equal brightness, and there are resolved systems with separations as small as 300~milliarcseconds. We find that a vast majority of wide binary systems that have been identified using various imaging techniques but are unresolved in \Gaia\ have RUWE values above 1.2, which supports the use of RUWE as an indicator of binarity \citep[e.g.][]{GaiaEDR3_Astrometry}.

This census is constructed to be expansive, but we err on the side of purity in terms of requiring objects to have some previous confirmation of youth. Our census has inherited the biases and incompleteness of its input catalogs. Many surveys of Taurus have focused near the clouds, introducing a location bias. Infrared and x-ray surveys are more likely to find Taurus members away from the clouds, but the former is biased towards younger, disk-bearing objects, and the latter is biased towards young solar-type stars that are active and bright enough in x-rays to be detected. This results in an incompleteness of the census away from the clouds and at ages older than canonical Taurus, qualifying the conclusions that can be drawn about older, distributed populations of Taurus. Large-scale, consistent searches for young ($\tau \lesssim 50$~Myr) stars without any initial phase space preference \citep[e.g.][]{Zari18,Kounkel19b,McBride20,Kerr21} are key in unraveling the picture of the greater Taurus region, and any as of yet unknown related populations around it.

\begin{deluxetable}{ll}
\tablecaption{Census of the Taurus-Auriga Complex without \Gaia\ EDR3 Sources\label{tab:nogaia}}
\tabletypesize{\footnotesize}
\tablehead{ \colhead{Column Label} & \colhead{Description } }
\startdata
Name & Common or other name\\
2MASSJ & 2MASS Point Source Catalog ID\\
R.A. & Right ascension J2000 (deg)\\
Decl. & Declination J2000 (deg)\\
Internal & In internal list\\
Kraus17 & In \citet{Kraus17}\\
Luhman18 & In \citet{Luhman18}\\
Esplin19 & In \citet{Esplin19}\\
Galli19 & In \citet{Galli19}\\
Jmag & $J$ magnitude (mag)\\
Jmag Ref & $J$ magnitude reference\\
Jmag Flag & Flag on $J$ magnitude\\
RV & Radial velocity (\kmsec)\\
e\_RV & Radial velocity uncertainty (\kmsec)\\
RV Ref & Radial velocity reference\\
SpT & Spectral type\\
e\_SpT & Spectral type uncertainty\\
SpT Ref & Spectral type reference\\
SpT Flag & Flag on spectral type\\
A\_mag & Extinction (mag)\\
e\_A\_mag & Extinction uncertainty (mag)\\
A\_mag band & Extinction band\\
A\_mag Ref & Extinction reference\\
Binary & Binary flag\\
Sys arc & Description of system architecture\\
Binary ref & Binary system reference\\
\enddata
\tablecomments{Numbered references for $J$ magnitude, radial velocity, spectral type, extinction, and binarity are the same as in Table~\ref{tab:census}, and listed in that table's comments.}
\end{deluxetable}

\subsection{Additional data compilation}\label{subsec:ancillary_data}

We have compiled additional data on the objects in our census needed for further analysis presented in this paper. These additional data include spectral types, extinctions, $J$ magnitudes, and radial velocities, which are listed for all objects with those measurements in Tables~\ref{tab:census} and \ref{tab:nogaia}.

A majority of the spectral types and extinctions were taken from the compilations of \citet{Kraus17}, \citet{Luhman18}, and \citet{Esplin19}, although the original reference is adopted where applicable. Spectral type and extinction values determined simultaneously from spectra were prioritized for adoption. Uncertainties on each are adopted from listed references. For spectral types from \citet{Esplin14}, \citet{Luhman18}, and \citet{Esplin19}, uncertainties of $\pm0.25$ and $\pm0.5$ sub-classes are used for values derived from optical and infrared spectra respectively, \citep[following][amongst others]{Luhman17}. For objects from \citet{Slesnick06b}, a spectral type uncertainty of 0.5 sub-classes is assumed from \citet{Slesnick06a}. The uncertainty on $A_V$ from \citet{Herczeg14} is assumed to be $0.25$~mag. For $A_V$ derived using photometry from \citet{Kraus17} we assumed an uncertainty of $0.3$~mag. For objects from \citet{Luhman09b}, \citet{Luhman17}, \citet{Luhman18}, and \citet{Esplin19}, we assume an uncertainty on $A_J$ of $0.1$~mag for extinctions derived from infrared spectra following \citet{Luhman17} and $0.14$~mag for extinctions derived from optical spectra and photometry following \citet{Luhman04} and \citet{Luhman03} respectively. \citet{Luhman17} says that the extinction derived from infrared spectra for objects with $A_J>2$ are more uncertain, so we adopt an uncertainty of $0.3$~mag for those objects. It is unclear if the same methodology is used across these papers, but these are the only available uncertainty estimates. We adopt an uncertainty of $0.45$~mag, the mean uncertainty in $A_V$ for objects with reported extinction errors, for objects without literature uncertainty values.

Regardless of the band of the adopted extinction, we have computed extinctions in $V$, $J$, and \Gaia\ EDR3 $G$ bands. To convert between $V$ and $J$, we assume an extinction coefficient of $R(J) = 0.72 \pm 0.01$ \citep{Yuan13}, which corresponds to $A_V/A_J = 4.31$ assuming $R(V) = 3.1$. Error is propagated from both the adopted extinction and $R(J)$. To convert to $A_G$, we adopt a temperature dependent relation for $A_G/A_V$ derived using theoretical spectral templates by Kraus et al. (in prep). We determine temperatures for the objects in our census using the adopted spectral types.

Nearly all of our adopted $J$ magnitudes were adopted from the 2MASS point source catalogue \citep{2MASS_catalog,2MASS_paper}. For select objects, we have adopted $J$ magnitudes presented in \citet{Esplin19} from other photometry sources.

We compiled radial velocities from various literature sources for our sample, prioritizing higher-precision literature values over \Gaia\ catalogue radial velocities. 14 objects have new radial velocities reported derived from observations with the Tull coud\'{e} optical echelle spectrograph on the 2.7-m Harlan J. Smith Telescope at McDonald Observatory \citep{Tull95}. The spectra are reduced with standard procedures implemented in \texttt{python}, which will be detailed in a future paper (Krolikowski et al., in prep), and radial velocities are extracted using broadening functions with theoretical model spectra \citep{saphires}. Roughly 70\% (415/587) of the objects with \Gaia\ counterparts have radial velocities, with a median uncertainty of 0.4~\kmsec. For objects without a literature radial velocity error, we adopt an uncertainty of 1~\kmsec.

For each object in our sample with \Gaia\ astrometry, we derived galactic Cartesian $XYZ$ positions by generating 100,000 \Gaia\ positions and parallaxes using the full \Gaia\ covariance matrix. We apply object-specific zero point corrections to the \Gaia\ parallaxes using the color, brightness, and location dependent relation from \citet{Lindegren20}\footnote{We use the \Gaia-provided \texttt{python} implementation of this relation that is found at \url{https://gitlab.com/icc-ub/public/gaiadr3_zeropoint}}. An analysis of eclipsing binaries by \citet{Stassun21} finds that these corrections from the \Gaia\ Collaboration are promising, and we include them for consistency with the \Gaia\ team despite their insignificance at the typical Taurus distance. We then convert to galactic $XYZ$ and adopt the median value. This allows us to compute individual $XYZ$ covariance matrices for all objects in our sample, which is important as $XYZ$ is highly covariant along the line-of-sight direction.

We combine the literature compiled radial velocities with \Gaia\ proper motions to compute full 3D kinematics of our sample. We compute galactic $UVW$ velocities for all objects with radial velocities by generating 100,000 \Gaia\ positions and astrometric values using the full \Gaia\ covariance matrix, again adopting the median value and computing individual kinematic covariance matrices. The median radial velocity uncertainty is roughly an order of magnitude larger than the reported projected proper motion uncertainty, which will again introduce a line-of-sight covariance to the $UVW$ velocity data. All \Gaia\ derived quantities are listed in Table~\ref{tab:census}.

\section{Finding Spatial Substructure in Taurus}\label{sec:substructure}

\subsection{Identifying subgroups with Gaussian mixture modeling}\label{subsec:gmm}

With this expansive Taurus census, we can robustly identify substructure in the region. We choose to identify clusters in galactic Cartesian coordinates, rather than with astrometric quantities, and excluding any kinematic information. We do not cluster in right ascension, declination, and parallax as they do not form an orthogonal coordinate system for a population of finite extent, and the covariance matrix of a Gaussian with non-zero spread in this space cannot encompass the curvature of the plane of the sky. As many objects do not have radial velocities, and thus do not have full galactic kinematics, we exclude kinematic information from the clustering and will analyze kinematics separately. This should not significantly affect clustering results, especially for well-defined tight spatial groups, and post-clustering kinematic analysis can provide further validation of our group assignments.

To find spatial clusters, we use a Gaussian mixture model \citep[hereafter GMM;][]{McLachlan00,Reynolds09}, which is a semi-supervised machine learning clustering model that assumes a data set can be fully described as a combination of multiple Gaussians. Each Gaussian component has its own mean and covariance matrix, allowing for elongated structures to be identified. Group membership is assigned probabilistically, accounting for covariances in the model Gaussians. This is a more accurate and robust model than traditional Euclidean distance based clustering algorithms, such as $k$-means clustering \citep{MacQueen67,Hartigan75} which partitions data into clusters by simply minimizing the square of the Euclidean distances between data points and their assigned cluster means, limiting model group shapes to be spherical in three dimensions\footnote{For a useful and simple visual explanation of $k$-means clustering, see \url{https://www.youtube.com/watch?v=4b5d3muPQmA}}.

Each component in a GMM has three defining parameters -- a mean vector, a covariance matrix, and a weight. The mean and covariances describe the location and shape of each Gaussian component, and the weight describes the relative prior probability for each component, which is essentially the fraction of objects in the fit data set belonging to each component. In a GMM, we can describe the probability distribution of a given data set $X$ as the sum of an ensemble of Gaussians:

\begin{equation}
    p(X) = \sum_{i=1}^{K} w_i~\mathcal{N}(X,\bm{\mu_i}, \Sigma_i)
\end{equation}

\noindent where $K$ is the number of components in the mixture model, $\mathcal{N}$ denotes the mulitvariate normal distribution, $\bm{\mu_i}$ and $\Sigma_i$ are the mean vector and covariance matrix of the $i$-th Gaussian component, and $w_i$ is the component weight. 

With this model, data points are assigned probabilities of being generated from each Gaussian component relative to the other components, rather than labels as in other clustering techniques. In this work, we assign objects the label corresponding to the highest component membership probability. We track the membership probabilities in each group for all objects in our analysis in case there are ambiguous group assignments.

\begin{figure}
\begin{center}
\includegraphics[width=\columnwidth]{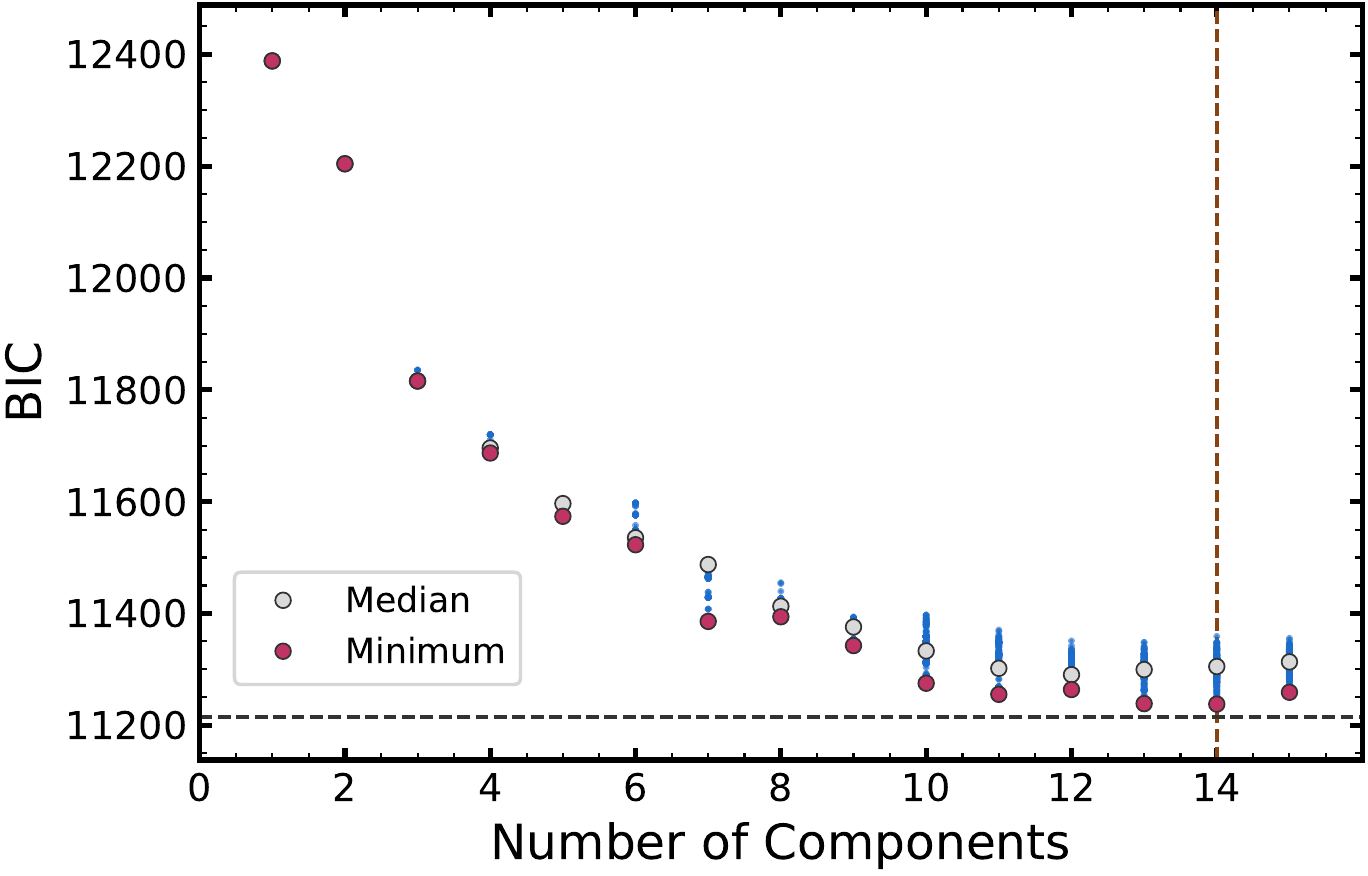}
\caption{Bayesian information criterion for the Gaussian mixture model fits as a function of number of components. For each number of components, 300 fits with different initialization states were run. All fit results are plotted as blue points, with the minimum BIC value (red circle) and median BIC value (gray circle) for each number of components plotted. The global minimum BIC (denoted by the vertical dashed line) occurs with 14 components. The BIC curve follows the expected trend of decreasing expeditiously as components are added to better the fit, before slowly rising again as the model becomes too complex. The BIC of the final adopted GMM fit is marked with the horizontal dashed line, and is smaller than the global minimum of the initial GMM fits.\label{fig:bic}}
\end{center}
\end{figure}

As we are searching in galactic $XYZ$ space, we can only define the GMM with the 528 objects in our census that have full \Gaia\ astrometric solutions. Additionally, we exclude 16 objects that have anomalously discrepant $XYZ$ locations with respect to the extended Taurus region, being too far away (with $X>225$~pc); none of these objects would be consistent with the broader Taurus region even accounting for parallax error. While these objects would likely just add additional outlier groups to the GMM fit, they would be extraneous and potentially bias the properties of actual groups related to Taurus on the outskirts of the region. These 16 excluded objects are denoted with an ``NM" (for non-member) in the Group Label column in Table~\ref{tab:census}.


We make no cuts to the sample based on the quality of an object's astrometric solution. Poor parallaxes could affect how the GMM finds clusters, particularly as the parallax error comprises nearly all of an object's $XYZ$ error and causes the values to be highly covariant in the line-of-sight direction. We address this later quantitatively with detailed fitting of well-defined groups accounting for each object's individual $XYZ$ covariance (see Section~\ref{subsec:mcmc}).

To perform the GMM fit to our data set we use the \texttt{GaussianMixture} module from the open source \texttt{scikit-learn} machine learning package in \texttt{python} \citep{sklearn}. We initialize our GMM fit using 100 iterations of $k$-means clustering. The number of Gaussian components that comprise the model are not fit parameters, but chosen. We calculate and compare the Bayesian information criterion (BIC) for all fits to determine the best fit. The BIC depends on both the output likelihood and complexity of a model, such that the preferred model will have a maximized likelihood while being penalized for the number of Gaussian components. Additionally, the GMM fit is slightly sensitive to the random seed for the $k$-means clustering initialization. 

To ensure we are finding the global BIC minimum, we run fits ranging from 1 to 15 components using 300 different random initial seeds to probe differences with the initialization, resulting in 4500 fits total (300 for each number of components). Figure~\ref{fig:bic} shows the BIC as a function of the number of GMM components for all fits, along with the minimum and median BIC for each number of components. The global minimum BIC fit has 14 components, which we will adopt as an initial set of groups for further investigation in Section~\ref{subsec:gmm_results}.

\subsection{Detailed subgroup spatial fitting to account for distance-related covariance}\label{subsec:mcmc}

One key drawback to the \texttt{scikit-learn} GMM routine is that it doesn't account for errors on the data themselves. In calculating the galactic $XYZ$ position of the objects in our census, the vast majority of measurement error is introduced by the parallax, which propagates to a strong covariance along the line of sight. For the Taurus region, the $X$ coordinate holds most of this parallax error, with the covariance strongest in the $XZ$ plane. From the 100,000 $XYZ$ samples we calculate for each of the objects in our census, we can compute an individual $XYZ$ covariance matrix for each object, which should be utilized in determining the physical properties of our GMM-found subgroups. 

Accounting for the data covariance is important to separate physical elongation from apparent elongation along the line of sight. To incorporate the data covariance information, we perform an MCMC fit of the individual GMM-found Taurus subgroups using \texttt{emcee} \citep{emcee}. We model each subgroup as a Gaussian defined by 9 parameters: 3 for the $XYZ$ mean position of the subgroup ($\mumod$) and 6 parameters defining the covariance of the model Gaussian ($\Sigma_{\rm mod}$).

A covariance matrix is a variance matrix that is rotated to introduce correlation between variables. We can decompose a covariance matrix into its eigenvalues and eigenvectors, which are the variances of the (un-rotated) variance matrix and rotation matrix respectively. The 6 parameters in our model defining the model covariance are 3 eigenvalues ($\bm{\lambda}$) and 3 angles ($\bm{\theta_\lambda}$). These parameters can be combined to generate a covariance matrix using the below equation, where $\mathcal{R}$ is the rotation matrix function and ${\rm diag}$ is the diagonal matrix function:

\begin{equation}
    \Sigma_{\rm model} = \mathcal{R}(\bm{\theta_\lambda}) \, {\rm diag}(\bm{\lambda}) \, \mathcal{R}(\bm{\theta_\lambda})^\intercal \label{eq:eigen}
\end{equation}

\noindent With this, we can generate random covariance matrices easily by generating 3 random angles and 3 random eigenvalues (greater than 0), without worrying about the need for directly generated random covariance matrices to be positive semi-definite.

We modify the data covariance matrices in two ways to account further for the under-reported nature of \Gaia\ astrometric errors. First, we multiply the data covariance matrices by the square of an object's RUWE. This will down-weight objects with worse astrometric solutions in the MCMC fit. If an object is an outlier but has large RUWE, it could very well be a bonafide member of the group but would bias the group's parameters from the GMM. Weighting by the RUWE will decrease the effect this outlier would have on the fit parameters. We also include an extra free parameter in our fit, $f_{\rm err}$, which is an error inflation term. While including the RUWE acts as an error inflation term specific to each object, $f_{\rm err}$ accounts for a uniform under-reporting of astrometric error, as is known to be the case in \Gaia. We do not allow for the case that astrometric errors are over-reported, so we require $f_{\rm err} \geq 1$.

\begin{figure}
\begin{center}
\includegraphics[width=\columnwidth]{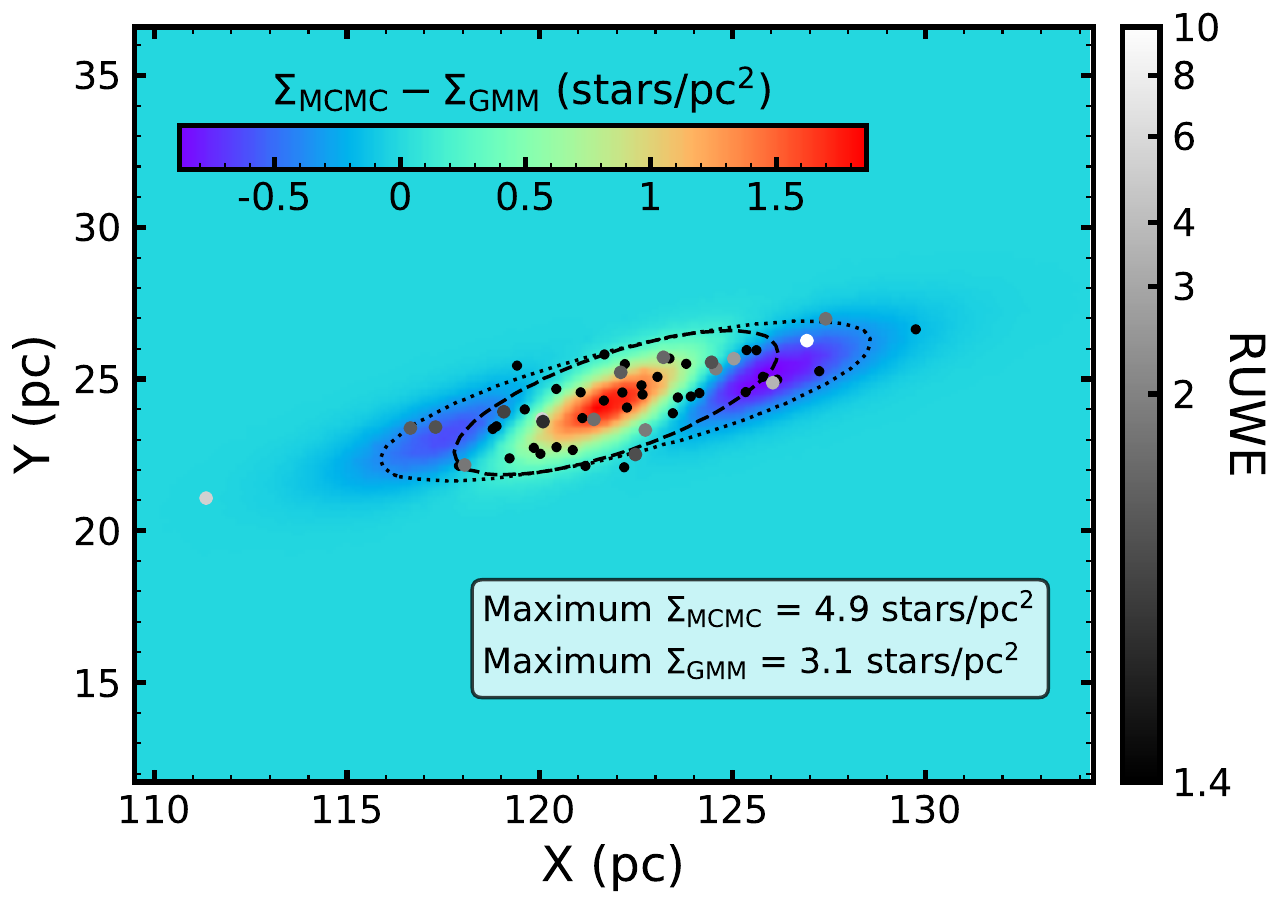}
\caption{Example result from the MCMC fit of Group \cTwo\ using the group specific error inflation term, showing the $XY$ slice. This slice is chosen as $X$ contains most of the line-of-sight dimension, and $Y$ is mostly in the plane of the sky. Data points are plotted colored with their RUWE value, and all points with RUWE lower than 1.4 (which denotes a good astrometric fit) are given the same darkest color and plotted as smaller circles. The points with worse RUWE values, especially those with the worst RUWE values, tend to be outliers along the line of sight. The 2$\sigma$ contours for the original GMM fit (dotted line) and MCMC fit (dashed line) are plotted, and the background image is the difference between the stellar areal density in this projected slice between the MCMC and GMM fits. The original GMM fit is larger than the MCMC fit, with virtually all elongation along the line of sight. The MCMC fit is more centrally concentrated than the GMM fit, highlighted by the roughly 60\% larger maximum stellar areal density for the MCMC Gaussian fit. \label{fig:mcmc_fit_example}}
\end{center}
\end{figure}

It is straightforward to incorporate the modified data covariance matrix into the likelihood calculation because covariance matrices add. Thus, the covariance matrix utilized in the final likelihood calculation for an individual data point, $\Sigma_{{\rm fit}, i}$, is

\begin{equation}
    \Sigma_{{\rm fit}, i} = \Sigma_{\rm model} + f_{\rm err}~{\rm RUWE}_{{\rm data},i}^2~\Sigma_{{\rm data}, i}
\end{equation}

\noindent The likelihood function over all $N$ data points is then

\begin{multline}
    \log \mathcal{L} = - \frac{1}{2} \sum_{i=1}^{N} (\bm{x_i} - \mumod)^\intercal \Sigma_{{\rm fit}, i}^{-1} (\bm{x_i} - \mumod) \\ + \log( 8 \pi^3 \, \det\Sigma_{{\rm fit}, i})
\end{multline}

In the MCMC fit, we do not allow the error inflation term to be less than 1, any of the model covariance matrix eigenvalues to be less than 0, and restrict the model covariance matrix angles to be between $-\pi$ and $\pi$. We also apply a Jeffreys prior on the covariance matrix eigenvalues:

\begin{equation}
    \log {\rm Prior}_{\bm{\lambda}} = - \frac{1}{2} \sum_{n=1}^{3} \log \lambda_n
\end{equation}

\noindent All parameters are initialized using the output for each particular subgroup from the GMM fit, and $f_{\rm err}$ is initialized uniformly between values of 1 and 6. We also run two different types of fits: one in which we fit each of the subgroups individually, and one in which we fit all of the subgroups together. In the latter, we use one global $f_{\rm err}$ term. This also allows us to compare the fit $f_{\rm err}$ values for each group individually to a global value, which will inform the degree to which each group is elongated along the line of sight, and any connection between error inflation and further substructure in a group.

Figure~\ref{fig:mcmc_fit_example} shows the results of the MCMC fit on one of our final adopted subgroups, Group \cTwo, using a group-specific $f_{\rm err}$ value. The group's spatial extent is shown using the probability density function of the fit Gaussian. The fit from the GMM is more extended than for the MCMC, which is smaller almost exclusively along the line-of-sight. The MCMC fit is also more centrally concentrated, meaning that the group is more compact than it appears in the GMM. Objects with RUWE above 1.4, which denotes a worse astrometric solution, are highlighted. Interestingly, these points tend to be on the outlying regions of the group along the line-of-sight. This confirms that the quality of the astrometric solution is significantly affecting the parallax, causing an apparent elongation along the line of sight.

\section{Spatial structure of the greater Taurus region}\label{sec:spatial_results}

\begin{figure*}
\begin{center}
\includegraphics[width=\textwidth]{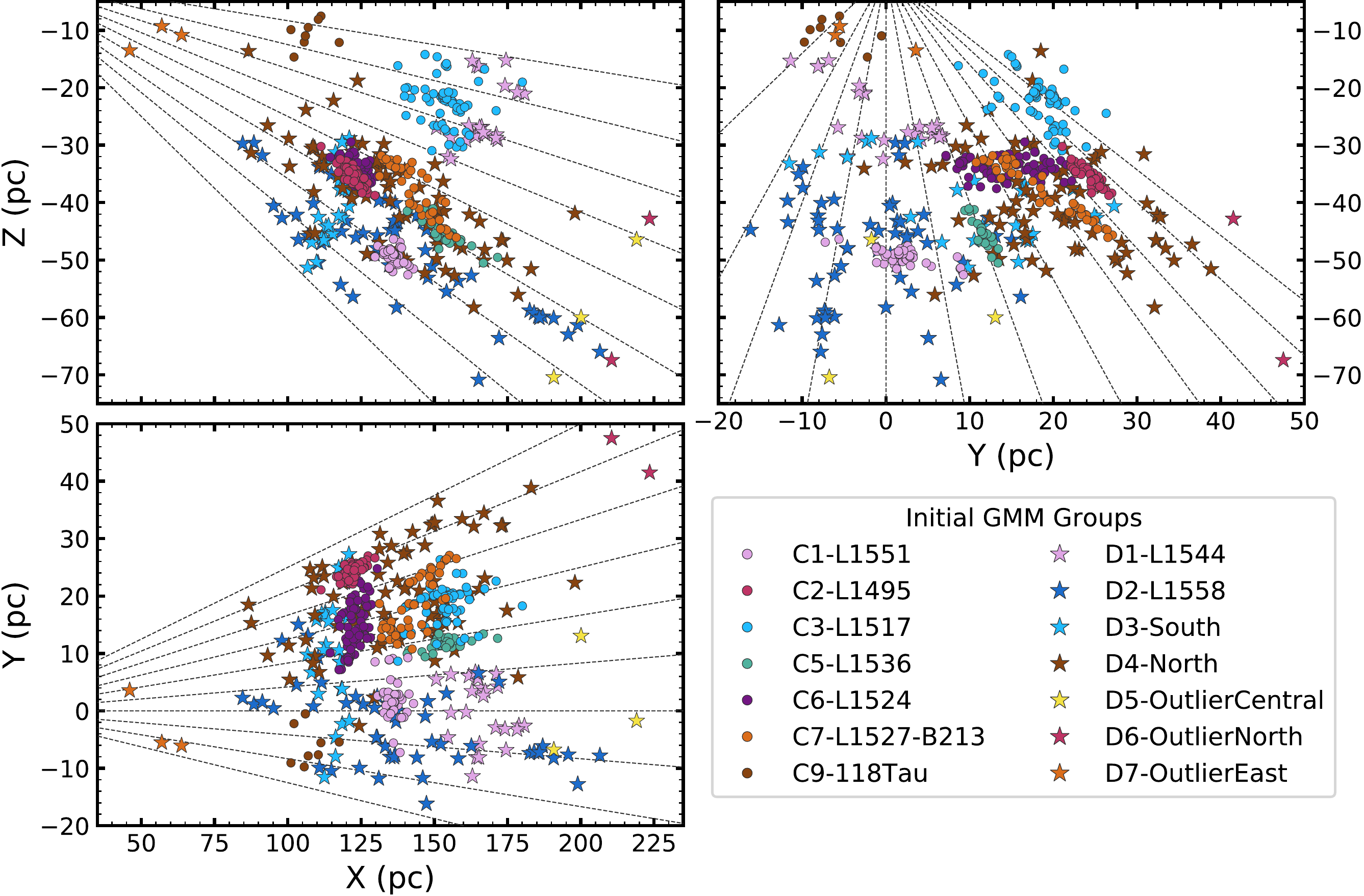}
\caption{Galactic positions of our census of the greater Taurus-Auriga region, presented in three slices through $XYZ$ space. Objects with \Gaia\ astrometry used in the Gaussian mixture model fit are plotted, with markers and colors denoting the group of which they are a member. The GMM fit used here is the original 14 component fit adopted. Note that the core group names skip numbers (C4, C8, and C10) which are additional groups adopted in the final GMM. The background dashed lines denote the line of sight directions across the region.\label{fig:xyz_min_gmm}}
\end{center}
\end{figure*}

\subsection{Characterizing the spatial subgroups of Taurus}\label{subsec:gmm_results}

The initial conditions and subsequent star formation history of the parent molecular clouds are imprinted in the substructure of the extended Taurus-Auriga region's stellar population. We use the GMM fit described in Section~\ref{subsec:gmm} to quantitatively identify subgroups in Taurus and assign group membership to our census, which will provide a basis with which to study the spatial, kinematic, and temporal structure of the broader Taurus region. In this section we analyze the output of the GMM fits, investigate simple properties of the identified subgroups, and finalize the subgroup membership assignments for our census.

Figure~\ref{fig:xyz_min_gmm} shows the $XYZ$ positions of our sample, plotted with colors and markers denoting their membership in groups identified in the initial GMM fit (note that not all 17 groups in the legend are used here, as further group subdivision will be explained below). There appear to be three types of groups identified by the GMM: 1) well-defined groups that are relatively confined in $XYZ$ and mostly reflect the core of the Taurus region (hereafter referred to as core groups and plotted as circles), 2) broad, sparser groups that have some apparent small-scale clustering but are found throughout $XYZ$ space (hereafter referred to as distributed groups and plotted as stars), and 3) groups with very few members ($N\leq3$) that are physically wide (hereafter referred to as outlier groups and plotted as stars).

\begin{figure}
\begin{center}
\includegraphics[width=\columnwidth]{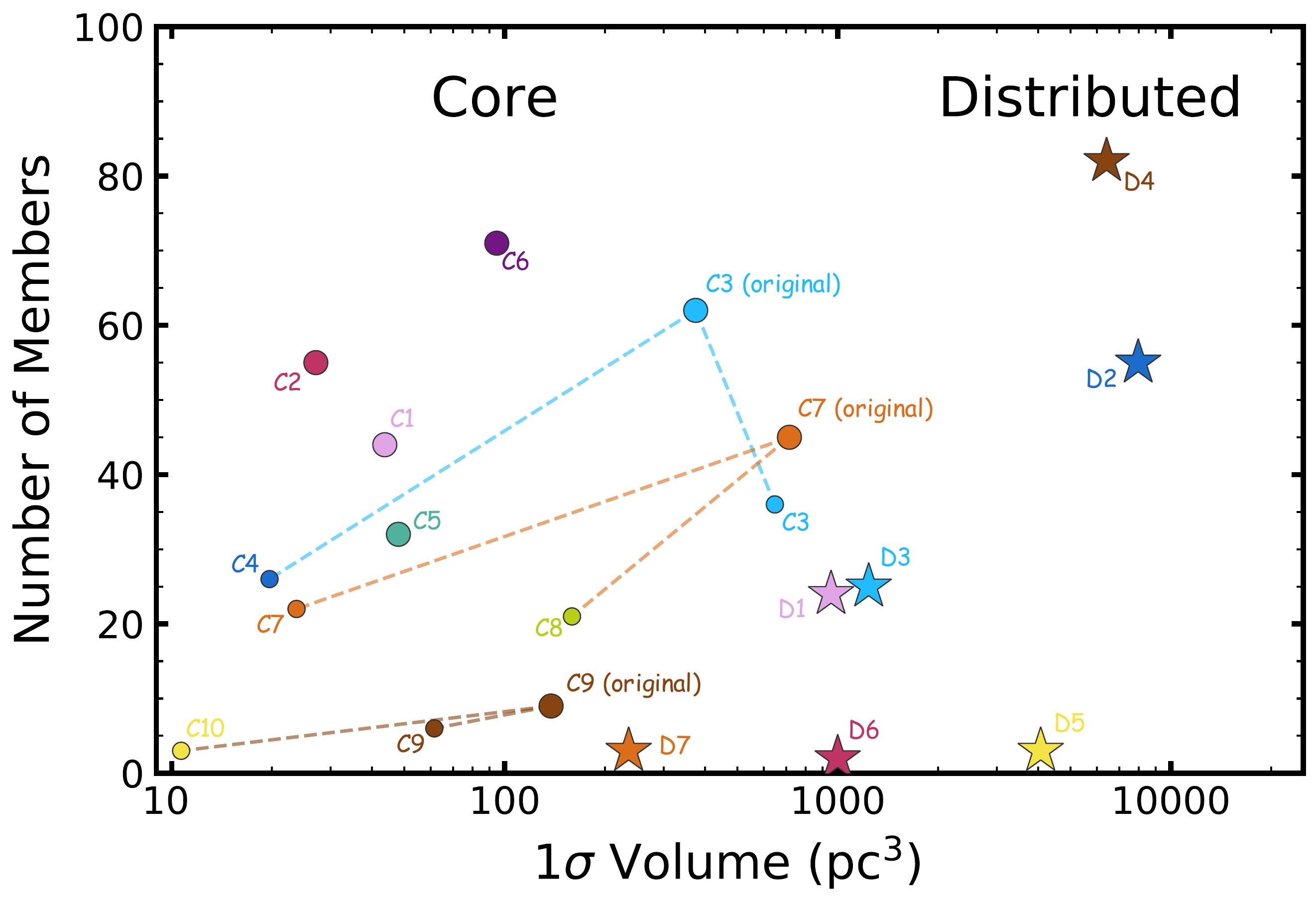}
\caption{Number density of the GMM-identified Taurus subgroups. The markers and colors retain their meaning from Figure~\ref{fig:xyz_min_gmm}. The number of group members is plotted against the group's 1$\sigma$ ellipsoid volume. The groups delineate themselves into three types: relatively small volume core groups, large distributed groups, and very lowly populated outlier groups. There are groups of intermediate density that we assign as either core or distributed, but could represent a transitional phase from compact to dispersed. Despite having comparable numbers of members, the distributed groups are substantially larger than core groups. The small plotted points are the subdivisions of core groups found with the subset GMM fits and adopted in the final 17 component GMM ensemble. Group \cSevenOG\ is split into two groups (the visually distinct loci Groups \cSeven\ and \cEight) that are core-like, and Group \cThreeOG\ is split into a centrally concentrated group (Group \cFour\ that is core-like) and its halo (Group \cThree\ that is of intermediate density).\label{fig:group_density}}
\end{center}
\end{figure}

We adopt a naming convention for the subgroups that conveys both the group type (core or distributed) and location. The names of core groups all start with ``C" and the names of distributed groups all start with ``D", followed by a number for distinction. Each group name has a description or cloud name appended to the end to convey its location within the Taurus region using just text. The outlier groups are named as distributed groups (D5-D7). There have been previous naming schemes for groups in Taurus, such as the roman numerals to distinguish groups on sky from \citet{Gomez93} and \citet{Luhman09b}, and the multiple different groupings using \Gaia\ in recent work \citep{Luhman18,Galli19,Roccatagliata20}. However, we do not adopt the roman numeral naming as the 2D grouping does not map exactly to the 3D grouping, nor any names from recent work as their groups are either too broad, too small, or use a significantly different Taurus census. The definitions of the groups in Taurus is ever evolving, leading to a difficulty in a single consistent naming scheme; we include a cloud name where possible to best orient the reader to the group locations.

This delineation can be seen in the number density of the groups, which is visualized in Figure~\ref{fig:group_density}. This plot shows the number of members in each group as a function of its $1\sigma$ ellipsoid volume. There appear to be four separate density regimes: groups with small volume and nearly an order of magnitude spread in membership count (Groups \cOne, \cTwo, \cFive, \cSix, \cNineOG), groups with many members and very large volume (Groups \dTwo, \dFour), groups in between these two regimes (Groups \cThreeOG, \cSevenOG, \dOne, \dThree), and groups with very few members across multiple orders of magnitude in size (Groups \dFive, \dSix, \dSeven). It is clear that the first of these regimes includes the core groups, the second includes the distributed groups, and the last is the outlier groups. For the groups in between, we consider Groups \cThreeOG\ and \cSevenOG\ to be core groups, and Groups \dOne\ and \dThree\ to be distributed groups. While Group \cThreeOG\ is extended, it appears to have a central concentration of stars with an extended halo, and Group \cSevenOG\ appears to have two physically distinct, compact loci. Groups \dOne\ and \dThree\ are considered distributed groups as they are sparser and less well-ordered. These groups with intermediate density may represent the transition from compact group at formation to a dispersing, evolved group; assigning them a core or distributed label is necessary for investigations of further subdivision. 

\begin{figure*}
\begin{center}
\includegraphics[width=\textwidth]{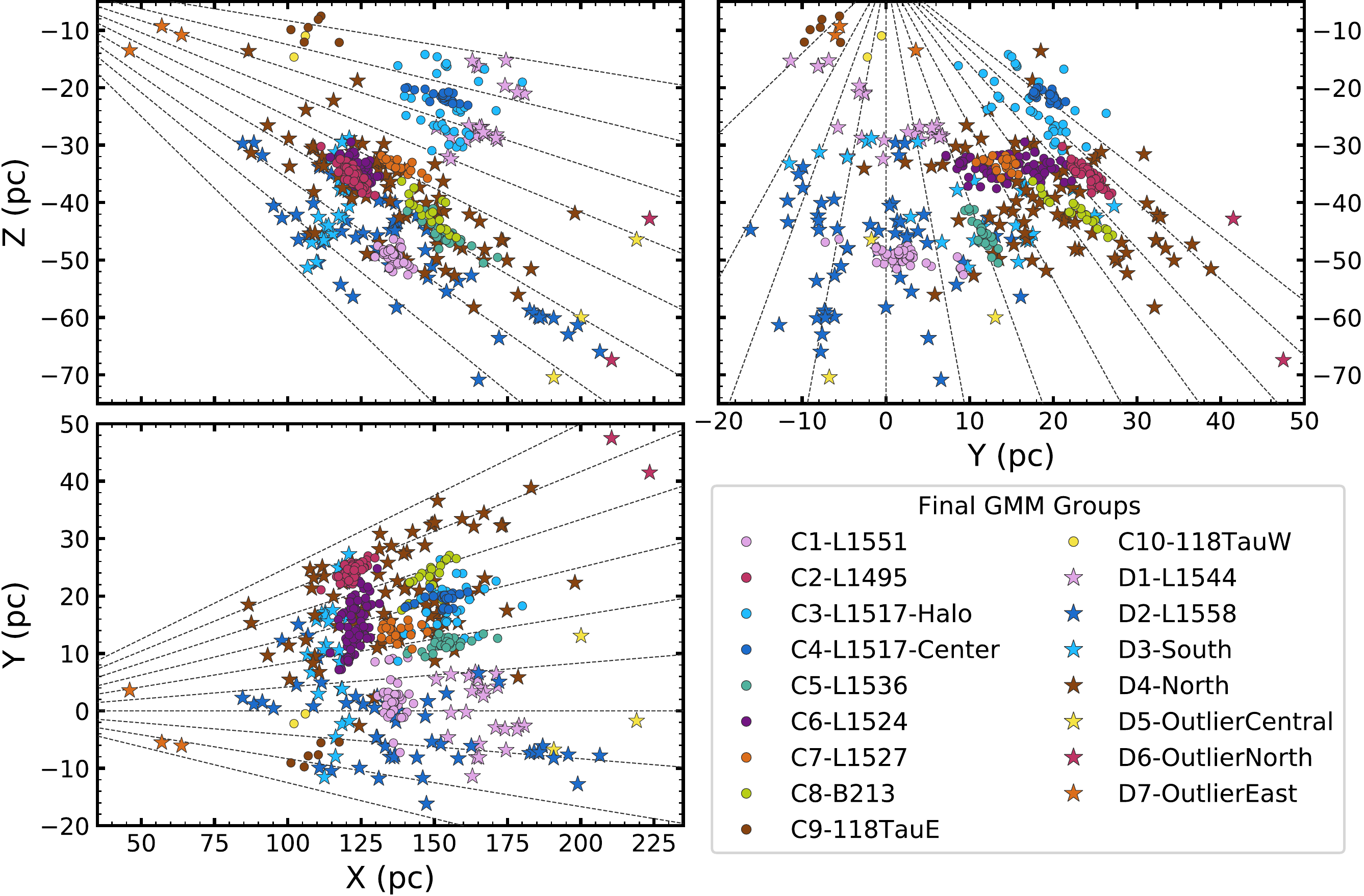}
\caption{Galactic positions of our census of the greater Taurus-Auriga region, presented in three slices through $XYZ$ space. The final adopted 17 groups from the combined subset Gaussian mixture models are shown. The background dashed lines denote the line of sight directions across the region. A 3D interactive version of this plot, showing the Taurus census in 3D galactic Cartesian space, is available to help visualize the region at \url{https://dkrolikowski.github.io/taurus_interactive_3d}. \label{fig:xyz_adopt}}
\end{center}
\end{figure*}

The core groups are elongated along the line-of-sight direction (denoted by the background dashed lines in Figure~\ref{fig:xyz_min_gmm}), particularly Groups \cTwo, \cThreeOG, \cFive, and both visually distinct stellar overdensities in Group \cSevenOG. We discuss this apparent elongation in Section~\ref{subsec:mcmc_results}. Groups \cOne\ and \cThree\ have halos of objects further from the mean of their respective groups (significantly more so for Group \cThree), but are still relatively compact and do not clearly overlap with any other group. Additionally, Group \cSevenOG\ appears to be two separate, spatially connected groups with visually distinct loci. The distributed groups, while having some apparent clustering, are mostly spread out across the full $XYZ$ space considered. 


To explore whether or not there is further substructure in the region not picked up by the original GMM (such as the apparent subdivision in Group \cSevenOG), we ran additional GMM fits on two subsets of our sample: the core and distributed groups separately. We include the outlier groups with the distributed groups, although there is no meaningful difference if they are included or not. We use the same methodology as with the initial GMM fit. 


For the core groups fit, the global minimum BIC is achieved with 11 components. We also look at the minimum BIC 10 component fit, which has virtually the same BIC as the global minimum fit. These two fits are nearly identical, splitting the two visually distinct loci of the original Group \cSevenOG\ (into the adopted Groups \cSeven\ and \cEight), distinguishing between the core and halo of the original Group \cThreeOG\ (split into the adopted Groups \cThree\ and \cFour), and splitting two members off of the original Group \cNineOG. Group \cNineOG\ was already small, but those two objects are distinct in $XYZ$. Groups \cNine\ and \cTen\ include four of the ten members of the 118 Tau association as defined in \citet{Gagne18}. These two groups are apparently associated with the 118 Tau association, and thus have adopted that name. The only difference between the two fits is a subdivision in Group \cSix\ in the 11 component fit that doesn't visually stand out. For this reason, and the nearly identical BIC values, we adopt the minimum BIC 10 component fit for the core groups, bringing our total number of core groups identified to 10.

For the distributed groups fit, the global minimum BIC is achieved with 6 components, although there is significant overlap in BIC values for fits with 4 to 8 components. Inspecting these fits, they preserve the original outlier groups, while adding one or two more small membership outlier groups and slightly rearranging the larger distributed groups. Notably, Group \dOne\ is kept exactly as it was in the original fit in all inspected subset fits. While there is apparent further substructure in these distributed groups, such as spatial correlations in all groups and overdensities in Groups \dOne\ and \dThree, the GMM struggles to pick up this clustering due to the broad and sparse nature of these groups. This could be explained by dispersion as clusters age, which would agree with the conclusion that these distributed groups are older than the core Taurus region \citep{Kraus17}. As such, we adopt the distributed and outlier group properties found with the original GMM.

For our final GMM group membership assignments, we adopt the core groups from the 10 component subset fit, and the distributed and outlier groups from the original 14 component fit. The BIC value of the adopted GMM ensemble is plotted as the horizontal dashed line in Figure~\ref{fig:bic}, and is smaller than the global minimum of the original set of fits, assuring that it is a reasonable fit in terms of membership assignment and complexity. Figure~\ref{fig:xyz_adopt} shows the $XYZ$ positions of our sample, along with a legend for the color and marker for each of these adopted Gaussian components (which will be used throughout the rest of the paper). Figure~\ref{fig:group_density} also shows the subdivisions of the core groups found in the adopted GMM ensemble. Various properties of the 17 spatial subgroups are given in Table~\ref{tab:groupprops}.

We recompute membership probabilities for all objects in the sample using these 17 Gaussian components. In the adopted fit, there are 317 members of the 10 core groups, 187 members of the 4 distributed groups, and 8 members of the 3 outlier groups. 


\subsection{\Gaia\ exaggerates the line of sight elongation of the subgroups in Taurus}\label{subsec:mcmc_results}

Almost all of the core groups have substantial spread in the line of sight direction, especially compared to their spread in the plane of the sky, with Groups \cTwo, \cFour, \cFive, and \cEight\ being the most elongated. Star formation often produces filamentary structures \citep{Hartmann02,Molinari10}, at least until those structures relax into clusters or disperse into the field \citep{Baumgardt07,deGrijs08}. As such, it is not unexpected that we would identify groups that are elongated along a particular axis. However, it would be highly coincidental for these groups to be truly elongated near exactly along the line of sight direction. To determine whether or not this line of sight elongation is real, we fit the core groups while accounting for the covariance in the astrometric data and the under-reporting of error in the \Gaia\ EDR3 catalogue. We do this by multiplying the data covariance matrices by both the square of their individual RUWE values, to capture the higher uncertainties for sources with poor fits, and by an error inflation term that is the same for all objects in a particular fit. We performed two types of MCMC fits: one on each group individually with their own error inflation term, and one on all of the core groups simultaneously with a global error inflation term. 

\begin{figure}
\begin{center}
\includegraphics[width=\columnwidth]{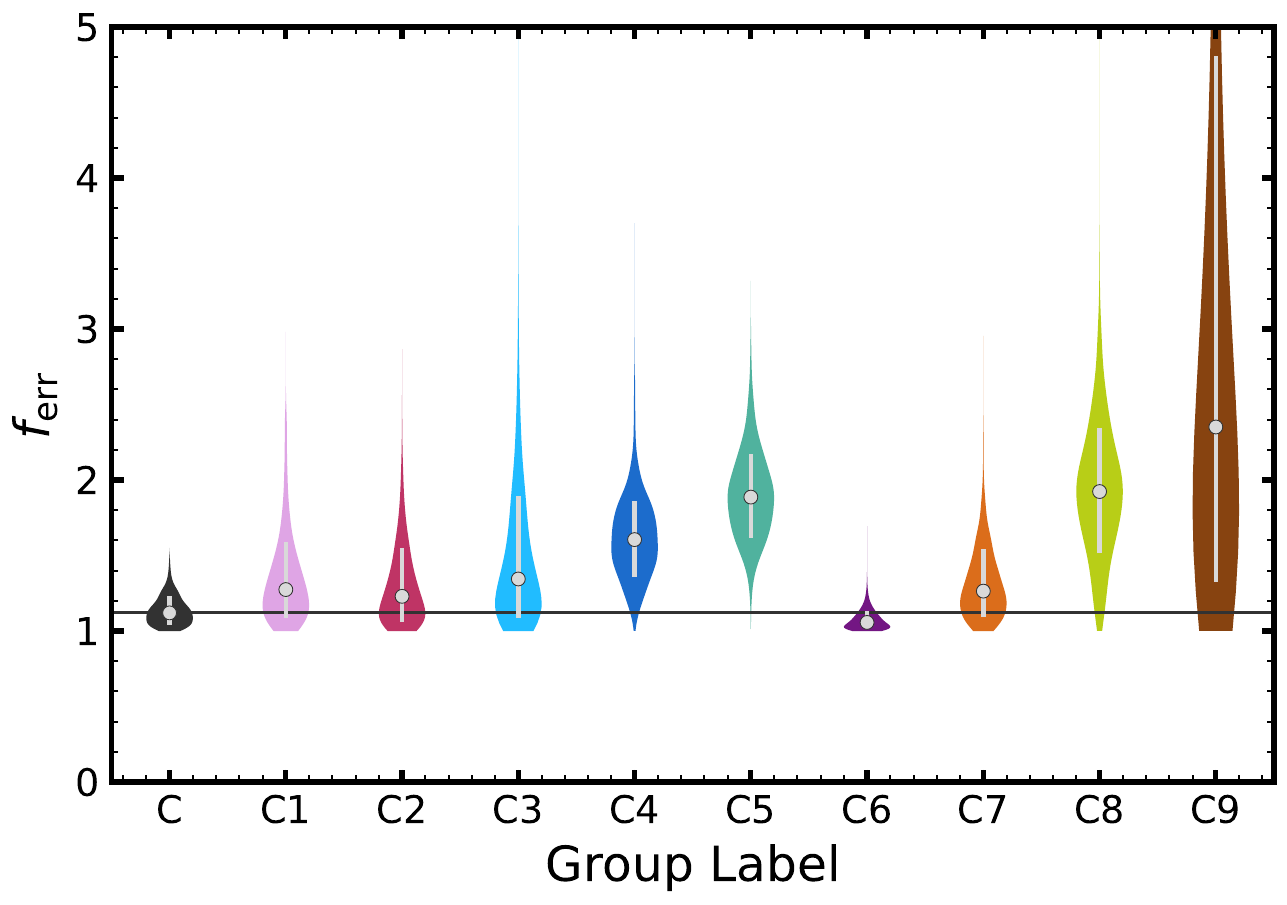}
\caption{Posterior distributions of $f_{\rm err}$ for the detailed MCMC fitting of the Taurus subgroups' spatial properties. The global fit, which has a single error term while fitting all of the core groups simultaneously, is plotted as group label C. The median value of the global fit is denoted by the horizontal black line.\label{fig:ferr_violin}}
\end{center}
\end{figure}

The global fit finds a median $f_{\rm err} = 1.12^{+0.11}_{-0.08}$, which means that \Gaia\ errors are under-reported by around 12\%. This agrees with the roughly 10\% error under-reporting found by \citet{GaiaDR2_RefFrame}, which is roughly the same as in EDR3 \citep{Fabricius20}. \citet{ElBadry21} looked at wide binaries in EDR3 and used their reported parallaxes and corresponding uncertainties to assess the degree to which the errors are underreported. They provide a relation to compute the expected parallax error inflation term for a single-star given its $G$ magnitude, which for the objects in the core groups produces a median $f_{\rm err} = 1.06^{+0.08}_{-0.03}$, in agreement with, but smaller than, the results of our global fit. The higher error inflation we find is likely attributable to additional noise introduced by the presence of unresolved binary companions, through orbital motion or photocenter jitter as the two stars vary in brightness independently over time.

Figure~\ref{fig:ferr_violin} shows the $f_{\rm err}$ posterior distributions from the MCMC group fits, including fits with a global error term and individual group-wise error terms. When the groups are fit individually, all groups except Group \cSix\ find an $f_{\rm err}$ larger than the global fit, ranging from 1.2 to 1.9. The $f_{\rm err}$ values for Groups \cFour, \cFive, and \cEight\ do not agree with the global $f_{\rm err}$. This is not surprising as these three groups are all highly elongated along the line of sight, confirming that the parallax error under-reporting is linked to this elongation. Group \cNine\ favors a much larger $f_{\rm err}$ (2.4) with a long tail towards even larger values because it only has 6 members, and thus the fit is ill-constrained and results in a large Gaussian fit as it has little structural knowledge to inform the fit. While we do not include them in the global fit, we did perform individual fits on the distributed groups. We exclude them as they are significantly broader than the core groups and do not have obvious elongation along the line-of-sight. Groups \dTwo, \dThree, and \dFour\ all have $f_{\rm err}$ smaller than the global fit, while Group \dOne\ finds a larger $f_{\rm err}$, due to it having a cluster of outlier points.

While the global fit provides insight into the catalogue-wide error inflation present in \Gaia\ EDR3, we adopt the individual fits for the purposes of analyzing the structure and size of groups. The individual fits are better at accounting for outliers in a group, in particular outliers that have relatively good RUWE values, as an increased $f_{\rm err}$ will be able to downweight those data in the fits when RUWE cannot. We interpret the individual $f_{\rm err}$ values not solely as direct measurements of the error inflation, but as a convolution of the error inflation term with properties of the individual groups. This includes further substructure, the frequency of undetected binary systems, and the spatial distribution of group members with poorly constrained astrometric solutions.

To examine the extent to which our subgroups are affected by error inflation along the line of sight, we compute the standard deviation for each group both along the line of sight and in the plane of the sky. To do this, we sample from the Gaussian function defining each group from both its GMM and MCMC fit. We then project the $XYZ$ positions of these samples onto the line of sight direction at the group's mean location. The norm of that projected vector is an object's line of sight coordinate, and the line of sight variance is then the variance of all group members' line of sight coordinates. We compute the plane of sky vector as the difference between the $XYZ$ position and line of sight vector, and again the plane of sky coordinate is the norm of this vector. This plane of sky projection is a one-dimensional displacement for a coordinate in a two-dimensional plane. To get the plane of sky variance, we find the displacement which encloses 39.3\% of the samples, which corresponds to the 1$\sigma$ distance for a 2D Gaussian.

\begin{figure*}
\begin{center}
\includegraphics[width=0.9\textwidth]{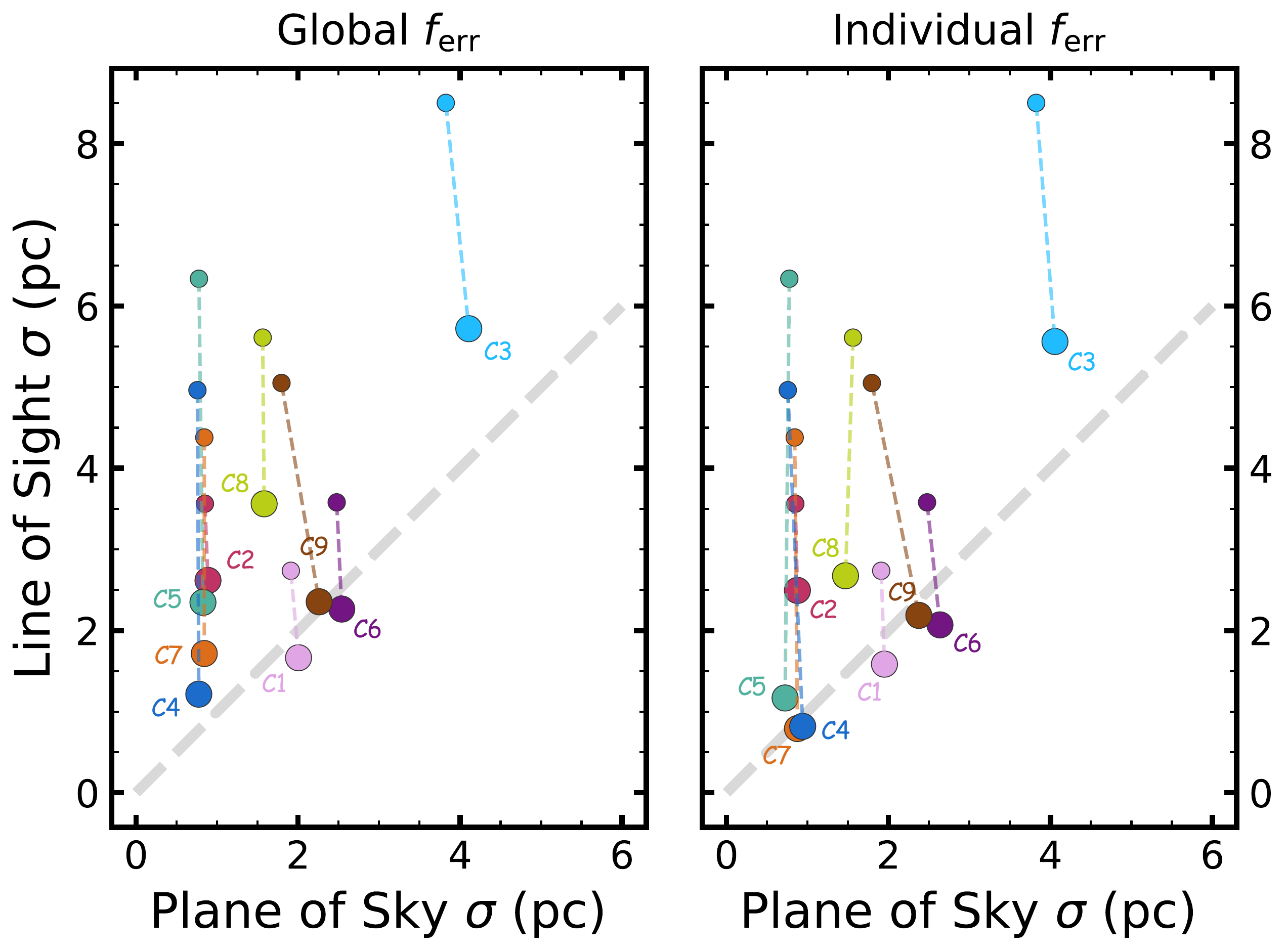}
\caption{Comparison of line-of-sight and plane-of-sky variances for each of the core Taurus subgroups. The colors retain their meaning from Figure~\ref{fig:xyz_min_gmm}. The small points are values from the adopted GMM fit and the large points are calculated from the output of the MCMC fit accounting for \Gaia\ errors, with the global fit on the left and individual group-wise fits on the right. All groups have smaller line of sight variances in the MCMC fit, showing that the elongation is an apparent effect from \Gaia\ finite parallax errors, reinforced by the negligible changes in the plane-of-sky variances between fits. \label{fig:spatial_sigmas}}
\end{center}
\end{figure*}

Figure~\ref{fig:spatial_sigmas} shows the line of sight and plane of sky standard deviations for both the GMM and MCMC fits of the core groups. The line of sight variance decreases for all groups, confirming that \Gaia\ error elongated the groups along the line of sight. This is reinforced by the fact that the plane of sky variance is virtually unchanged in all groups. All groups also fall closer to the one-to-one line with the MCMC fit results compared to the GMM fit results, meaning they are closer to equal aspect ratio (although they may have structure in the plane of the sky). The distributed groups do not have substantial change in either their line-of-sight or plane-of-sky variances, confirming that their large spatial extents are physical and not apparent. 

To validate the shrinking along the line-of-sight, we additionally ran fits arbitrarily setting $f_{\rm err} = 1$ to ensure the inflation term isn't solely responsible for shrinking the spatial extent of these groups. The fits with a free $f_{\rm err}$ are slightly smaller, particularly along the line-of-sight, which is expected since the additional uncertainty should allow for a smaller Gaussian to encompass a group's members. Otherwise, the fit Gaussians are largely similar in shape and orientation, and this assures that the bulk of the line-of-sight shrinking is due to the distribution of the lowest-uncertainty objects being least elongated, as quantified through the RUWE-corrected data covariances. 

As we are accounting for the parallax effect, we can be confident any remaining evidence of structure is real and not apparent. Groups \cTwo\ and \cEight\ are still larger in the line-of-sight than plane of the sky, by roughly a factor of 2. Group \cEight\ does not have many members, which could be affecting the fit quality, but Group \cTwo\ is very well populated. This indicates that Group \cTwo\ and Group \cEight\ are in reality elongated along the line of sight. These groups are associated with the clouds L1495 and B213, which have been suggested to have clumpy or layered structure in the line of sight velocity projection by \citet{Hacar16}. Our work adds that the cloud is likely spatially structured in the line of sight as well, and that Groups \cTwo\ and \cEight\ are the result of star formation along filaments in the cloud that have collapsed earlier than the present gas.

\movetabledown=1.7in
\begin{rotatetable}
\begin{deluxetable*}{lccccccc}
\tabletypesize{\footnotesize}
\tablecaption{Taurus Group Properties\label{tab:groupprops}}
\tablehead{ \colhead{Group} & \colhead{$(\alpha, \delta, \pi)$\tablenotemark{a}} & \colhead{Distance} & \colhead{$(X, Y, Z)$} & \colhead{$(U, V, W)$} & \colhead{$N_{\rm mem}$} & \colhead{Age\tablenotemark{b}} & \colhead{Marker\tablenotemark{c}}\\
\colhead{} & \colhead{(deg, deg, mas)} & \colhead{pc} & \colhead{pc} & \colhead{\kmsec} & \colhead{} & \colhead{Myr} & \colhead{} }
\startdata
\cOne & (68.083, 18.0493, 6.909) & 144.74 & (136.03, 2.09, -49.41) & (15.75, -15.22, -7.38) & 44 & $1.73^{+0.24}_{-0.2}$ & \LARGE \textcolor[HTML]{DFA5E5}{$\bullet$} \\
\cTwo & (64.3793, 28.1684, 7.72) & 129.53 & (122.23, 24.27, -35.36) & (16.22, -12.28, -10.72) & 55 & $1.34^{+0.19}_{-0.18}$ & \LARGE \textcolor[HTML]{BF3465}{$\bullet$}\\
\cThree & (74.089, 29.5632, 6.356) & 157.33 & (154.56, 17.98, -23.22) & (16.02, -14.64, -11.09) & 36 & $2.34^{+0.52}_{-0.38}$ & \LARGE \textcolor[HTML]{21BCFF}{$\bullet$}\\
\cFour & (74.1644, 30.3172, 6.444) & 155.19 & (152.48, 19.25, -21.53) & (14.8, -14.84, -10.47) & 26 & $2.53^{+0.97}_{-0.5}$ & \LARGE \textcolor[HTML]{1C6CCC}{$\bullet$}\\
\cFive & (68.7759, 22.8293, 6.22) & 160.76 & (153.76, 11.66, -45.46) & (16.47, -14.23, -6.9) & 32 & $2.01^{+0.32}_{-0.29}$ & \LARGE \textcolor[HTML]{50B29E}{$\bullet$}\\
\cSix & (68.0159, 25.2741, 7.788) & 128.40 & (122.95, 14.53, -34.04) & (15.78, -11.53, -9.49) & 72 & $1.57^{+0.18}_{-0.17}$ & \LARGE \textcolor[HTML]{731683}{$\bullet$}\\
\cSeven & (70.2087, 25.7439, 7.051) & 141.81 & (137.16, 13.92, -33.25) & (15.33, -11.61, -9.5) & 22 & $2.59^{+0.74}_{-0.78}$ & \LARGE \textcolor[HTML]{DB6D1B}{$\bullet$}\\
\cEight & (66.5258, 26.5957, 6.413) & 155.94 & (148.54, 22.7, -41.69) & (17.53, -13.08, -6.77) & 21 & $3.09^{+0.92}_{-0.72}$ & \LARGE \textcolor[HTML]{B8CE17}{$\bullet$}\\
\cNine & (83.8523, 22.845, 9.129) & 109.54 & (108.83, -7.54, -9.87) & (13.3, -18.94, -8.79) & 6 & $6.13^{+2.63}_{-1.64}$ & \LARGE \textcolor[HTML]{874310}{$\bullet$}\\
\cTen & (80.4945, 24.8425, 9.485) & 105.43 & (104.72, -1.08, -12.2) & (13.25, -18.73, -8.71) & 3 & $41.46^{+34.97}_{-24.36}$ & \LARGE \textcolor[HTML]{F4E345}{$\bullet$}\\
\dOne & (78.3837, 24.1845, 5.949) & 168.09 & (166.2, -0.12, -25.14) & (19.13, -14.15, -8.81) & 24 & $3.40^{+0.9}_{-0.72}$ & \large \textcolor[HTML]{DFA5E5}{$\bigstar$}\\
\dTwo & (69.7134, 17.3824, 6.784) & 147.40 & (139.22, -1.7, -48.37) & (15.43, -14.03, -8.64) & 55 & $3.27^{+0.42}_{-0.36}$ & \large \textcolor[HTML]{1C6CCC}{$\bigstar$}\\
\dThree & (66.2198, 22.0251, 8.138) & 122.89 & (115.87, 10.75, -39.49) & (14.01, -6.81, -9.52) & 25 & $6.22^{+1.43}_{-1.68}$ & \large \textcolor[HTML]{21BCFF}{$\bigstar$}\\
\dFour & (66.6245, 25.7332, 7.0) & 142.86 & (135.99, 19.01, -39.42) & (16.17, -12.66, -9.17) & 83 & $2.49^{+0.35}_{-0.34}$ & \large \textcolor[HTML]{874310}{$\bigstar$}\\
\dFive & (71.5097, 20.0666, 4.724) & 211.71 & (203.33, 1.53, -58.95) & (16.12, -8.21, -5.14) & 3 & $17.66^{+22.62}_{-9.61}$ & \large \textcolor[HTML]{F4e345}{$\bigstar$}\\
\dSix & (65.6635, 29.6987, 4.38) & 228.30 & (217.03, 44.49, -55.12) & (16.82, -7.78, -4.42) & 2 & $4.71^{+8.13}_{-2.28}$ & \large \textcolor[HTML]{BF3465}{$\bigstar$}\\
\dSeven & (77.5648, 20.4195, 17.612) & 56.78 & (55.6, -2.66, -11.2) & (10.04, -5.95, -9.64) & 3 & $23.23^{+34.53}_{-9.93}$ & \large \textcolor[HTML]{DB6D1B}{$\bigstar$}\\
\enddata
\tablenotetext{a}{\Gaia\ EDR3, J2016}
\tablenotetext{b}{Uncertainties are from the 16th and 84th percentiles of the age posteriors.}
\tablenotetext{c}{The marker and color used for each group in the plots in this paper.}
\end{deluxetable*}
\end{rotatetable}
\clearpage

\section{Kinematic structure in the extended Taurus region}\label{sec:kinematics_results}

In addition to high-precision galactic positions for our Taurus census, \Gaia\ also provides an unprecedented view of the kinematics of the region through a combination of its high precision proper motions and literature radial velocities (RVs). Kinematics are important to further verify the coherence of a spatially-identified collection of stars, in particular for those that are dispersed. The difference between an object's velocity and its parent group can be used to further vet membership. Perhaps most importantly, the velocity structure of a complex region like Taurus can be used to back out the timeline of its star forming event, and can give insight into how the resultant stellar population is tied to the initial conditions of the molecular cloud from which it was born \citep[e.g. Larson's Law;][]{Larson81,Solomon87}. 


\begin{figure*}
\begin{center}
\includegraphics[width=\textwidth]{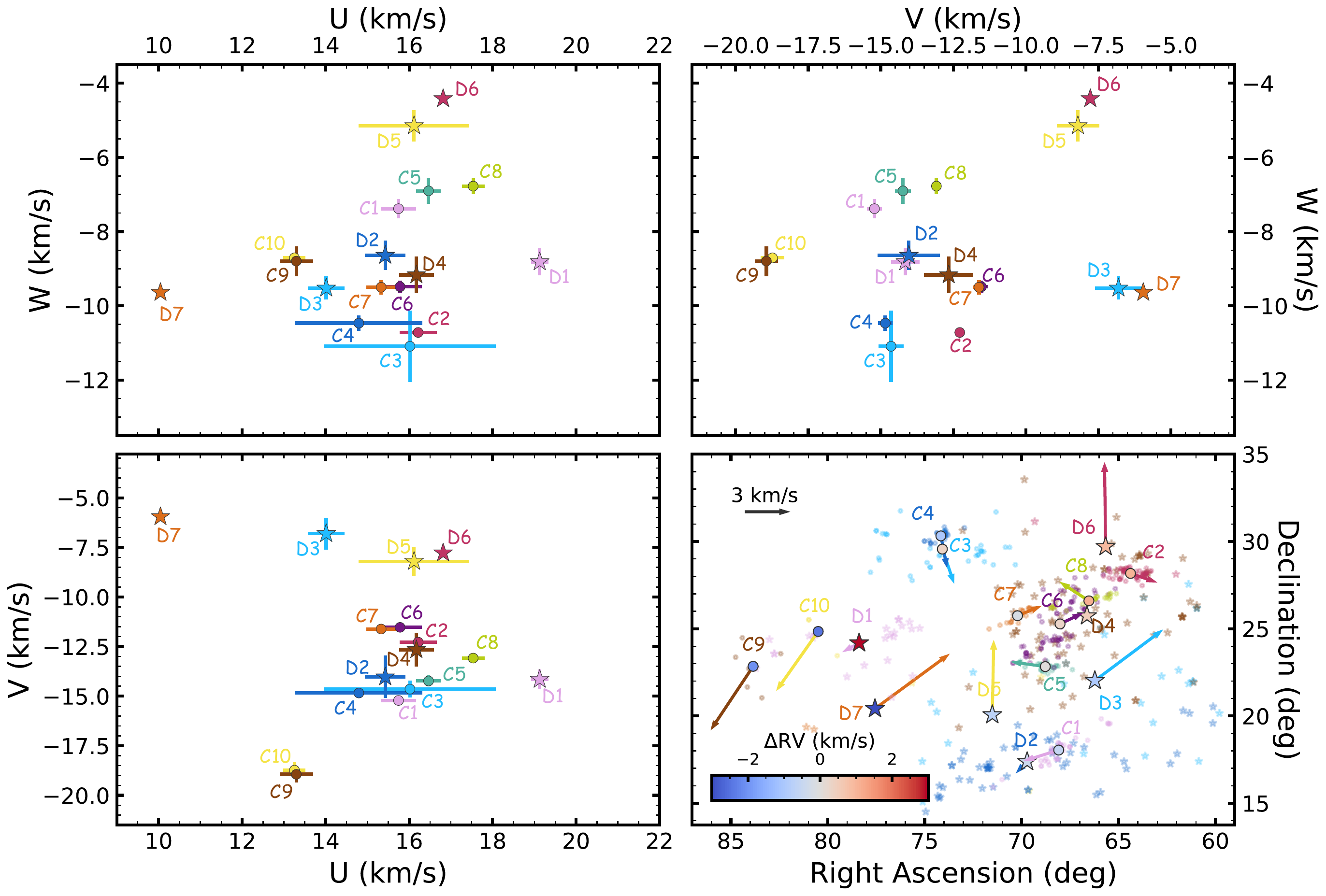}
 \caption{Bulk motions of the GMM-identified Taurus subgroups. Colors and markers retain their meaning from Figure~\ref{fig:xyz_min_gmm}, although the marker fill color in the bottom right panel represents an RV difference shown in the colorbar. Upper left, upper right, and bottom left panels: the bulk galactic $UVW$ motions of the subgroups. The group velocities plotted are the median values of group members with full 6D phase space information. An RV quality cut is used: objects with $\sigma_{\rm RV} \geq 0.5$~\kmsec\ and objects with RVs from LAMOST are excluded. Groups \dSix\ and \dSeven\ only have one member each that pass the RV quality cut, so the velocity plotted is of single objects. The error bars are the standard deviation in the mean in each dimension to visualize uncertainty in the group's bulk velocity vector. 
 Bottom right panel: the difference between the bulk motion of the subgroups and the average motion of the entire region, $(U,V,W) = (15.78, -13.08, -8.81)$~\kmsec, projected onto each subgroup's location. The large markers are the sky projected spatial centers of the subgroups, and are colored by the difference between their mean RV and the region RV at their location. The arrows are the difference between the proper motions of each subgroup and the region, converted to a tangential velocity. We show the difference to highlight the relative motions of the subgroups, such as the coherence between Groups \cOne, \cFive, and \cEight. This set of groups appears to be moving in the opposite direction of other core Groups \cTwo, \cSix, and \cSeven. There also is a tenuous positive RV gradient from the southeast to the northwest.\label{fig:uvw_medians}}
\end{center}
\end{figure*}

\subsection{The bulk motions of Taurus subgroups}\label{subsec:bulk_kinematics}

If the groups we have identified in Taurus are from distinct formation events, we expect them to have small internal velocity dispersion, with larger differences between the groups. The relative velocities of the groups can also shed light on the star formation history of the region, such as whether or not some subsets of the groups formed together. Figure~\ref{fig:uvw_medians} shows the bulk galactic $UVW$ velocities for all subgroups, which are presented in Table~\ref{tab:groupprops}, along with the standard deviation in their mean velocity vectors. We only include objects that have RV errors less than $0.5$~\kmsec, and exclude objects that have compiled RVs from LAMOST, as they are subject to systematic uncertainties larger than our criterion \citep{Guo15,Luo15,Zhong15,Kiman19,Luo19,Zhong19}. Two of the outlier groups (\dSix\ and \dSeven) have only one member with a good RV each, while the third outlier group only has two. Therefore, we do not analyze their kinematics further.

The core groups, except for Groups \cNine\ and \cTen\ which will be discussed later, have bulk motions within 4~\kmsec\ of each other in all three kinematic dimensions. There is coherence present, linking Groups \cOne, \cFive, and \cEight\ together, and Groups \cTwo, \cThree, \cFour, \cSix, and \cSeven\ together, although the latter set of groups has more dispersion. While there is some distinction in $V$, nearly all of this separation occurs in $W$. This parallels the spatial clustering: Groups \cOne, \cFive, and \cEight\ are all more distant and lower in $Z$ than the other set of groups. Groups \cSix\ and \cSeven\ are particularly close in both kinematics and spatial position. The kinematics of Groups \cThree\ and \cFour\ agree, which is consistent with their spatial coincidence. They also have a dearth of precise RVs compared to other groups, leading to a larger uncertainty in their bulk motions, which is most readily apparent in their large extent in $U$. Group \cThree\ also has a larger velocity dispersion than Group \cFour, which is consistent with being a dispersed halo of objects around Group \cFour. The RV uncertainties dominate the errors on the bulk kinematics of these groups. The $U$ velocity carries most of the RV information, and for all core groups has the largest error of the three velocity components.

The two core groups that do not follow this categorization are Groups \cNine\ and \cTen, which are significantly removed from where the main Taurus groups sit in velocity space, particularly in $V$. This is not surprising, as these two groups are significantly removed from the main Taurus group spatially, are similar to each other spatially, and older. Their kinematics however are not discrepant with being linked to Taurus. On average, they are separated from the other core groups by roughly 50 pc in space and 6.3 \kmsec\ in velocity. This is in line with expectations from Larson's Law ($\sim5-7$~\kmsec, depending on the formulation), and is similar in scale to spatial and kinematic substructure seen in the Sco-Cen association \citep{Rizzuto11}. It is possible that this small group was formed in a slightly earlier epoch of star formation linked to the recent and ongoing star formation in the main Taurus region. If Groups \cNine\ and \cTen\ are a part of the greater Taurus ecosystem, ongoing and future searches for young stars in the region between them and the core groups should have velocities that fall in between their bulk motions.

The distributed groups are less kinematically homogeneous than the core groups, and do not all reside in the same general location in velocity space as the core groups. The distributed groups have larger typical velocity errors than the core groups. This is particularly evident in their larger $V$ and $W$ error bars, which are much less affected by the RV error, which dominates the uncertainty budget. The larger velocity errors are indicative of a larger velocity dispersion, which can be explained if these groups are dispersed collections of coeval, potentially older stars. Groups \dTwo\ and \dFour\ are both kinematically close to the core groups, sitting in roughly the center of the core groups' $UVW$ distribution. This bolsters the idea that these two groups comprise stars either born in an earlier epoch of star formation near the Taurus region that have dispersed or that have formed in isolation concurrently with the core groups. In fact, while Groups \dTwo\ and \dFour\ are spatially distinct from each other (especially in $Y$), they can visually be joined together to create one large group of objects interspersed throughout Taurus with roughly the same kinematics. This distinction in Y could be due to these groups originating from two main sites of ongoing star formation: near Group \cOne\ and clouds L1551 and L1558 for Group \dTwo, and in the main canonically-defined Taurus complex near Groups \cTwo, \cEight, and \cSix\ and clouds L1495, B213, L1521, and L1524 for Group \dFour. This is analogous to Groups \cOne, \cFive, and \cEight, which are spread across 10s of parsec in the $Y$ direction, yet are kinematically coherent.

Groups \dOne\ and \dThree\ are more distinct, with Group \dOne\ removed from the other groups in $U$ and Group \dThree\ removed in $V$. Neither of these groups are particularly spatially distinct, although Group \dOne\ is separated from the other distributed groups. Group \dThree\ directly overlaps Groups \dTwo\ and \dFour. Even without kinematics, the GMM identified the snake-like structure of Group \dThree\ crossing through Groups \dTwo\ and \dFour. This structure is further highlighted by the fact that Group \dThree's plane of sky spatial standard deviation is nearly three times larger than that along the line-of-sight. Coupled with its distinct kinematics, Group \dThree\ is a distinct subgroup of Taurus with a filamentary or planar structure unlike nearly all of the other identified groups.

\begin{figure*}
\begin{center}
\includegraphics[width=\textwidth]{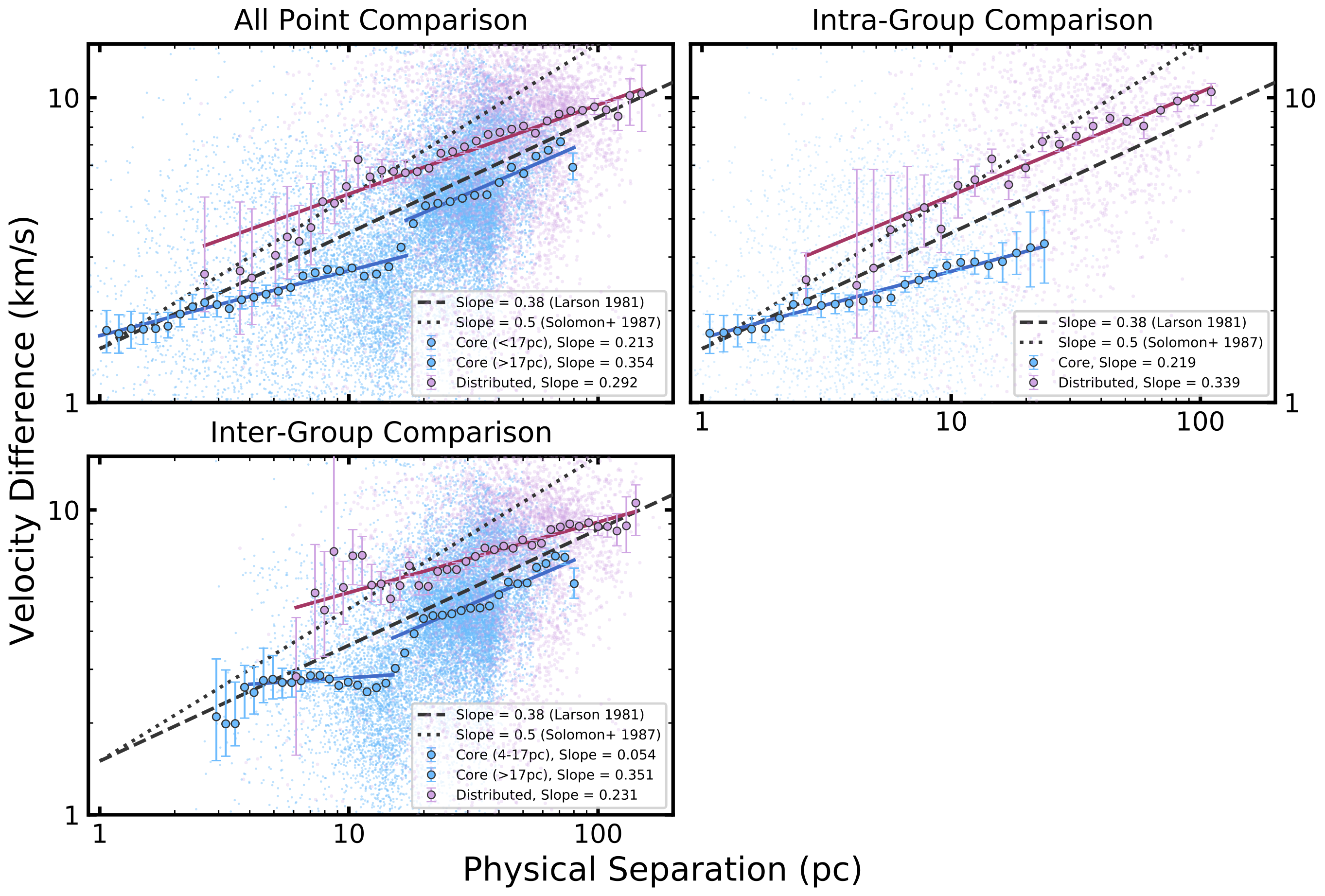}
\caption{Velocity structure functions (VSF) for the separate core and distributed group subsets of our census. Upper left panel: The VSF is calculated for each type of group (core or distributed) using all objects in a group of that type. Upper right panel: The VSF is calculated by comparing objects only to other members of their specific subgroup, and then combined with all other subgroups in their respective group type (core or distributed). Bottom left panel: The VSF is calculated by comparing objects only to other objects that are not members of their specific subgroup. The small background points are the VSFs of the data themselves. The large points are median bootstrapped VSF values from 1000 different instances of the corresponding VSF, where each instance uses samples of the position and velocity for each object to account for uncertainty. The solid lines are the adopted power law fits for each VSF, with broken power laws adopted for the all-point and inter-group VSFs of the core groups. The black dashed and dotted lines show power laws with slopes from previous studies of the velocity dispersion in molecular clouds for comparison \citep[][respectively]{Larson81,Solomon87}. For the core groups, the VSF is dominated by intra-group comparisons on scales smaller than roughly 17~pc. It is dominated by inter-group comparisons at larger scale, which has higher dispersion and a steeper slope. There is evidence in the inter-core-group VSF for a large energy injection at a roughly 17~pc scale. Since the distributed groups are large and dispersed, there is no strong separation-scale delineation between intra- and inter-group comparisons. The fit for the intra-distributed-group VSF agrees with the original formulation of Larson's Law. However, all slopes found are shallower than more recent studies that find a power of 0.5 for the velocity dispersion in molecular clouds. As stars are only perfect tracers of their birth environments at formation, the relations here likely reflect early dynamical evolution from the structure of the cores from which these stars were formed.\label{fig:vsf}}
\end{center}
\end{figure*}

\subsection{Testing turbulence with the velocity structure function of Taurus}\label{subsec:turbulence}

Stars form in molecular clouds that are turbulent, which is important in the star formation process as it leads to and regulates fragmentation, helps set the initial mass function, and is the seed of a stellar association's velocity field. Larson's Law is an empirical relation that relates the size of molecular clouds to their velocity dispersion, which is set by the turbulent power spectrum \citep{Larson81}. The velocity structure of our sample can directly test if Larson's Law applies to the groups we have found in Taurus in describing the turbulent nature of the region, and how much evolution occurs in the velocity structure after stars form and decouple from the gas. To do this we compute the velocity structure function \citep[VSF;][]{McKee07} of our sample, calculating the pairwise velocity difference between objects as a function of their 3D spatial separation. We calculate VSFs for multiple subsets of our data, such as using only objects that are members of core groups or distributed groups together. 

Computing VSFs using just the positions and velocities of the objects themselves does not account for the uncertainties on their 6D phase space coordinate, largely from parallax and RV error. There are also covariances introduced as different VSF points can have the same start or end point, so uncertainty in one position can introduce a covariance on the velocity structure information of $N-1$ baselines across the $N$-body system. To account for the uncertainty and covariance introduced in the VSF computation, we generate 1000 samples of galactic position and kinematics for each object included in a VSF. We draw these realizations of position and velocity from their \Gaia\ astrometric solution and covariance matrix, and our literature compiled RV and corresponding error. To avoid inflation of the VSF by observational uncertainties, we exclude all objects that do not meet the RV quality condition described in Section~\ref{subsec:bulk_kinematics}. With this, we produce 1000 different instances of the VSF encapsulating the underlying uncertainty in the velocity structure of the Taurus stellar population.

Figure~\ref{fig:vsf} shows three sets of VSFs for the core and distributed groups separately: one in which all objects in each type of group (core or distributed) are compared to each other (``all point comparison"), one in which objects are only compared to other objects in their specific subgroup (``intra-group comparison"), and one in which objects are only compared to other core or distributed objects not in their specific subgroup (``inter-group comparison"). There are differences between each of the subsets of the sample that are highlighted by the fits shown, particularly between the core and distributed group subsets.

The VSF data follow power laws, as expected from Larson's Law and formation in a turbulent molecular cloud, and we fit their slope and intercept to quantify the turbulent structure of the region. For a particular VSF, we fit each of the 1000 VSF instances separately, and then adopt values and uncertainties from the resulting slope and amplitude parameter distributions. We exclude all VSF points with 3D spatial separations below 1 parsec, as that is the typical distance uncertainty, and any points below that will be effectively smeared out and hold no precise structure information. 

For the fit, we first bin each VSF into equally-spaced logarithmic bins in 3D spatial separation. The bin edges are computed using the \texttt{astropy} implementation of Knuth binning, which is a Bayesian estimation of the optimal binning that is robust to the underlying data distribution \citep{Knuth06}. We bin using the corresponding data value-only VSFs (the adopted phase space coordinates of each included object, no sampling); we use the same bins on all VSF instances. Within each bin of 3D spatial separation, we calculate the median velocity difference, and fit the resulting binned VSF with a power law function. We exclude any bins with fewer than 5 data points in them, to reduce biasing the fit from outlier points with small number statistics, particularly at the smallest and largest scale separations. We adopt the median of the 1000 fits as the VSF parameter values and the 16th to 84th percentile range as the $1\sigma$ parameter uncertainties. For some of the fits, as will be described below, we fit multiple power laws over different spatial separation regimes.

The distributed group VSFs are accurately described with single, well-behaved power laws for each of the 3 cases. The slope for the intra-group VSF is larger than the slope for the all-point VSF, which in turn is larger than the inter-group VSF. Both the intra- and inter-group VSF slopes, however, agree with the all-point VSF within $1\sigma$. As the distributed groups are large and sparse, there isn't a clear spatial scale delineation between the inter- and intra-group comparisons. The intra-group VSF power law slope is similar to the original formulation of Larson's Law ($\sigma_v = 1.1~L^{0.38}$;~\citealt{Larson81}). Assuming that the stars have retained memory of the parent molecular cloud's turbulent structure, this would indicate that the groups formed from a cloud with subsonic turbulence. The inter-group VSF has a shallower slope, which could result from the different distributed groups forming from different clumps in the larger molecular cloud, such that they have retained the velocity structure within their clump but did not inherit any velocity structure beyond that level.

The core group VSFs are significantly more complex. The all-point VSF is a clear broken power law, with a change in amplitude and slope at roughly 17~pc. The smaller-scale regime is dominated by the intra-group VSF and the larger-scale regime is dominated by the inter-group VSF. To determine if there is further underlying structure, or if there is simply a change of power law at that scale, we must inspect the separate inter- and intra-group VSFs. 

The intra-core-group VSF is a steady, single power law across its entire separation range, and has a nearly identical slope to the small-scale regime of the all-point VSF. The inter-core-group VSF is quite different from the intra-core-group VSF, with two clear regimes: a nearly flat VSF below 17~pc and a power law above 17~pc with a nearly identical slope to the all-point large-scale VSF. The inter-core-group VSF has a large jump in velocity dispersion suggesting a significant energy injection into the entire core group region at a $\sim17$~pc scale. There are few data points in the intra-core-group VSF above that scale, so it is hard to conclude if the energy injection is evident there as well. The inter-core-group core VSF is essentially flat at 3~\kmsec\ between 4 and 17 parsecs, which means that even where individual core groups are closer to each other, there is no velocity structure between their members. 

We also interpret the difference between the inter- and intra-core-group VSFs at small scales to further validate the core group assignments. If these group identifications were spurious, we would expect the two VSFs to be identical at small scales. However, we see that the velocity difference between objects is smaller when they are members of the same group than when they are not at a given separation below 10~pc. This is another confirmation that group membership is meaningful with respect to possible association with the other groups. There are many fewer objects at the smallest scales in the inter-distributed-group VSF which makes it hard to determine if this is true for the distributed groups as well, although we expect that to be the case (and do see some apparent flattening at the smallest scales).


There are two major ways in which the VSF of a collection of stars can evolve over time, one which steepens the relation and one which makes it shallower. Objects at smaller separations, while having lower velocity dispersion, will move fractionally further away from each other than objects that are at large separations. This will result in the objects at small-scales ``catching up" to the objects at large-scales, steepening the slope of the VSF. On the other hand, groups of stars can dynamically relax for as long as they remain bound, losing memory of the initial velocity structure of the region and producing a shallower slope (asymptotically approaching a flat VSF) where there is no pairwise spatial and kinematic correlation. 

This process occurs over the relaxation timescale, which is given as:

\begin{equation}\label{eq:trelax}
    t_{\rm relax} = \frac{N}{6 \log{\frac{N}{2}}}~\frac{R}{\sigma_v}
\end{equation}

\noindent where $N$ is the number of stars in the group, $R$ is the radius of the group (taken as the half-mass radius), and $\sigma_v$ is the velocity dispersion; the second fractional term in the equation is the crossing time. For our groups, we can only estimate a relaxation timescale because their membership may not be complete and our number count is a lower limit. Also, relaxation may begin before molecular cloud material is expelled, leading to a deeper potential well with faster typical stellar velocities. We assume that our groups are mass-mixed, and that the half-mass radius is equivalent to the half-count radius. To calculate the half-count radius, we take 10000 samples using each group's mean and covariance matrix output from the MCMC spatial fits using an individual $f_{\rm err}$ term. We then determine the 1D distance at which half of the samples are enclosed. We take the velocity dispersion to be the standard deviation in 1D velocity separation for objects in a group that pass the RV quality cut from \ref{subsec:bulk_kinematics}.

We can interpret the VSFs presented here in the context of these two processes to draw conclusions about the nature of the groups. The slope of the intra-core-group VSF is much shallower than that of the intra-distributed-group VSF. The core groups are much more compact than the distributed groups, with calculated relaxation times ranging from 0.5 to 3 Myr for the core groups and 4.3 to 9 Myr for the distributed groups. With these short relaxation times, combined with the typically assumed Taurus age of less than a few million years, we conclude that the individual core groups have begun to dynamically relax while sufficient molecular cloud material remained to keep them bound, resulting in a shallower VSF slope. If the distributed groups are young and formed as they are seen now -- sparsely and across a large spatial extent -- they have not had enough time to relax, and will more closely retain the velocity structure of the parent molecular cloud. If they are older, evolved, dispersing equivalents of the core groups, they may have already dynamically relaxed to some extent, before gas expulsion unbound the groups and their dispersal led to a steeper VSF slope.

It is harder to apply these processes to the inter-group VSFs, particularly for the core groups as they are distributed through the region in spatial clusters, and thus have no well-defined crossing time, unlike the distributed groups which are regularly spaced throughout the region. The inter-core-group VSF's steeper slope is in agreement with the intra-distributed-group VSF's slope, implying that the original velocity structure of the cloud has not been lost on the larger scale between groups. It is unclear why the inter-distributed-group VSF has a shallower slope, but it perhaps is an indication that the distributed groups are less related to each other than the core groups are to themselves. If they formed at different times, unlike the core groups, we don't expect them to behave as though they formed from the same molecular cloud.

The slope of the velocity structure function of stars in Taurus is shallower than the currently accepted typical Larson's Law slope of $0.5$ from \citet{Solomon87}. \citet{Qian18} measured the velocity structure function of molecular cores in Taurus identified in \citet{Qian12} over projected separation scales from $1-10$~pc. They find that the molecular core VSF follows a power law with a slope between $0.5$ and $1/3$, with a steeper slope below 3~pc and a shallower slope above 5~pc. This is in agreement with the roughly $1/3$ slope we find for the VSFs we believe trace the initial spectrum of the stellar population, although we find much higher velocity dispersions at all separations than in the molecular core VSF. Stars inherit the velocity structure of their parent molecular cores, except exhibiting higher velocity dispersion, perhaps from early dynamical interactions when the stars decoupled from their parent cloud. The agreement between the VSFs of these two different types of objects implies that the turbulent power spectrum in Taurus may be shallower than other molecular clouds.


\section{Star Formation History of the Taurus Region}\label{sec:ages}

While the spatial and kinematic properties of stellar groups can provide information about the structure of a star forming event, ages are crucial to tease out the star formation history. Age estimates of young stellar populations are difficult because they are acutely sensitive to observational uncertainties from extinction, accretion, distance, and binarity, and require mostly complete censuses to achieve high precision. Kinematics can be used to estimate ages via traceback analysis, which has proved challenging in the past \citep{Blaauw64,Brown97,Mamajek05,Pecaut12}, but may now be feasible with \Gaia\ astrometry \citep{Chronostar,Wright20}. Traceback requires a well vetted, complete census, high precision velocities, and the assumption that the star forming event occurred in a compact region \citep{Wright20}, which is not the case for at least the full ensemble of Taurus members. Therefore, to construct the star formation history of Taurus, we must assess ages directly using the well-established method of isochrone fitting.

Stellar models at young ages have few observational calibrations and are subject to substantial uncertainties \citep{Pecaut12,Rizzuto16,Rizzuto20}. There are also astrophysical sources of uncertainty, such as accretion history that may add intrinsic scatter \citep{Baraffe17} and unresolved binaries \citep{Sullivan21}. Altogether, these issues make it hard to derive absolute ages from photometry and isochrones. However, photometric ages can be measured consistently, and then used to assess the relative ages of groups in a star forming region to construct a star formation timeline. With these ages we can assess the star formation history of the region by looking at inter-group age gradients, such as with size and location, and at the age of individual groups, for the presence of an age spread. 

\begin{figure*}
\begin{center}
\includegraphics[]{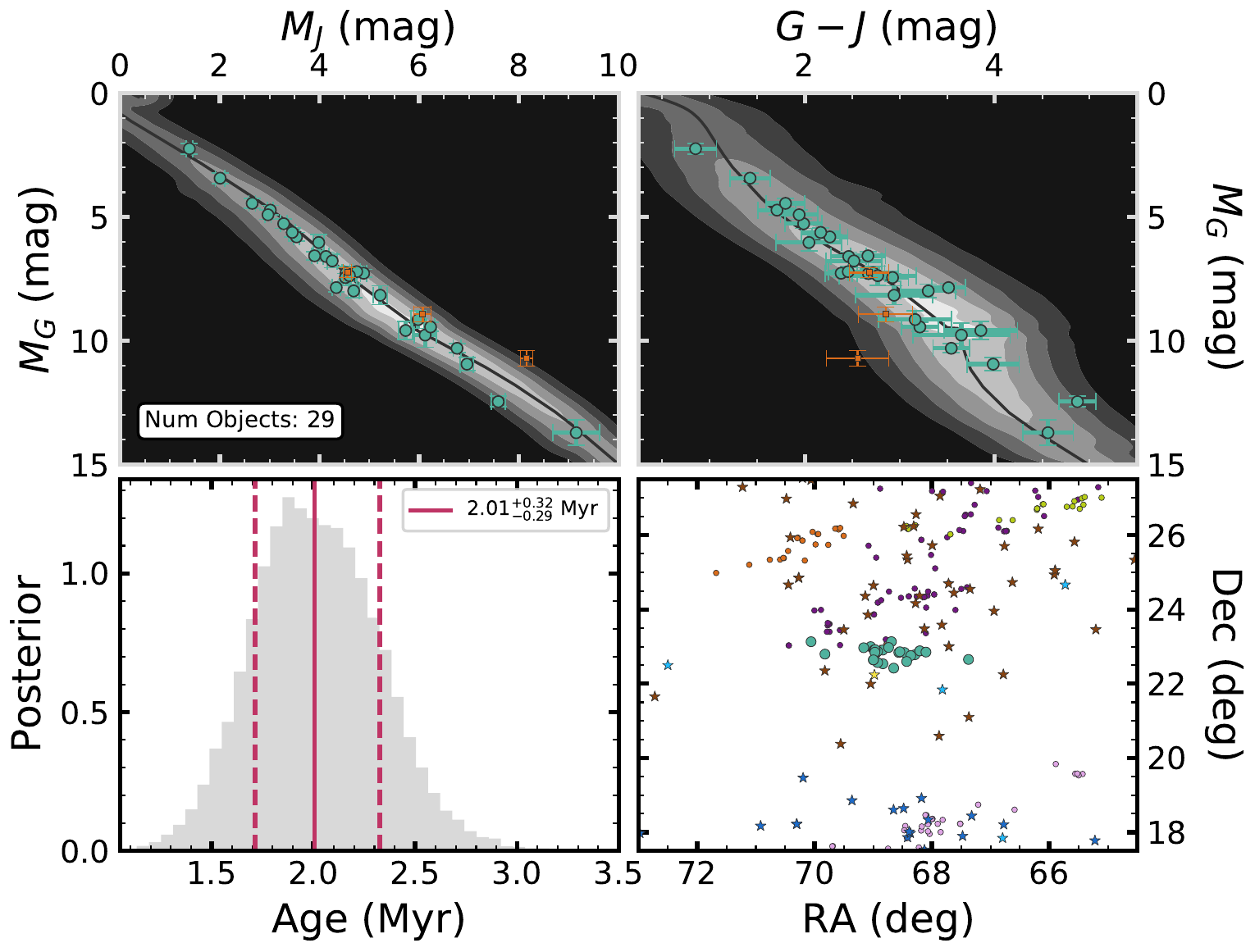}
 \caption{Example results of the Bayesian age inference code for Group \cFive. The upper two panels show the magnitude-magnitude (the space where the fit is done) and color-magnitude diagrams for the group. The round teal points are objects included in the fit, and the orange square points are objects that have been excluded (in this case either from extinction that is too large or Class I identification). The KDE of the best-fit median age is plotted in the back, along with the best-fit isochrone as the dark line. The bottom left panel shows the posterior distribution of ages from the fit, with the 16th and 84th percentiles marked. The bottom right shows the sky position plot highlighting the location of Group \cFive\ (plotted as the large points).\label{fig:age_code_example}}
\end{center}
\end{figure*}

\subsection{Ages from Bayesian isochrone fitting}\label{subsec:agecode}

To derive ages for the Taurus subgroups, we construct a Bayesian framework for calculating ages from isochrones for a collection of stars. To calculate the likelihood of a star (or system) falling at a specific HR diagram or magnitude-magnitude diagram position, we compute the underlying probability density function of a synthesized stellar population of a given age. We draw each synthetic star's magnitude from an isochrone and broaden the ensemble's magnitude-magnitude sequence using a kernel density estimator (KDE) to generate a smooth distribution. With the KDE likelihood, we compute a posterior distribution of age. We finally explore the posterior space using a Markov Chain Monte Carlo with \texttt{emcee} \citep{emcee} to find the best-fit age and its uncertainty. 

For this work, we use \Gaia\ EDR3 $G$ and 2MASS $J$ magnitudes. We do not use \Gaia\ $G_{\rm BP}$ or $G_{\rm RP}$ magnitudes as they both can be particularly sensitive to the presence of binaries, and have less flux and fewer scans than $G$. Due to the common presence of disks, we choose 2MASS $J$ over $K$ as the disk excess will add additional $K$-band flux above the photospheric level, which we do see bias our fits towards younger ages if $K$ magnitudes are used.

We first generate the ensemble of model stars of the given age with masses drawn from a three-power Kroupa IMF, as presented in Equation 2 of \citet{Kroupa02}. We then interpolate  masses and age on an isochrone grid to get $G$ and $J$ magnitudes for each model star. Here we use a combination of the BHAC15 and MIST isochrone grids \citep{Baraffe15,Dotter16,Choi16}. The MIST grid is only computed down to $0.1~{\rm M_\odot}$, while the BHAC15 grid is computed down to $0.01~{\rm M_\odot}$. We adopt the MIST grid above $0.1~{\rm M_\odot}$ and the BHAC15 grid below. There is a slight discontinuity at the transition between the two models ($0.1~{\rm M_\odot}$), but it is smaller than the typical measurement uncertainty. The MIST grid has been updated to include the \Gaia\ EDR3 bandpasses, while the BHAC15 grid hasn't, but the differences between the MIST DR2 and EDR3 magnitudes are much smaller than typical uncertainties. We therefore have chosen to use the BHAC15 grid's DR2 colors in interpreting the observed EDR3 photometry.

\begin{deluxetable*}{ccl}
\tabletypesize{\normalsize}
\tablecaption{Mass-dependent multiplicity fraction\label{tab:binaryprops}}
\tablehead{ \colhead{Masses (${\rm M_\odot}$)} & \colhead{MF} & \colhead{Reference} }
\startdata
$<0.15$ & 0.33 & \citet{Kraus05}\\
$0.15-0.4$ & 0.35 & \citet{Kraus12}\\
$0.4-1.5$ & 0.45 & \citet{Tokovinin20}, \citet{Raghavan10}\\
$>1.5$ & 0.7 & \citet{DeRosa14}\\
\enddata
\end{deluxetable*}

We have also implemented a binary population synthesis to account for unresolved multiples. Following \citet{Sullivan21}, we assign a binarity flag to each model star using a mass-dependent multiplicity fraction shown in Table~\ref{tab:binaryprops}. We then draw a separation for each binary system using the mass-dependent log normal distributions described in \citet{Sullivan21}. This prescription covers objects with masses between 0.1 and 2~M$_\odot$; we adopt the 0.1~M$_\odot$ distribution for objects with masses below this range and the 2~M$_\odot$ distribution for objects more massive than this range. We then draw a mass ratio from a power law distribution whose index is mass-dependent. We use the same piecewise linear interpolation described in \citet{Sullivan21} to generate these mass dependent power law distributions, with one exception: we account for the dichotomy seen in the mass ratio distribution at high primary masses. \citet{DeRosa14} finds that A-star binaries with separations smaller than 125~AU have a more uniform mass ratio distribution than wider separation systems. We therefore replace the highest mass point in the piecewise interpolation with a power law index value of -0.5 for systems with separations below 125~AU. For the binary systems, we compute magnitudes for each component from the interpolated isochrone grid.

Depending on the binary system's separation, we then combine the individual object magnitudes to get an unresolved system magnitude. We assume a common distance of 145~pc to convert all physical separations into angular separations, and use different angular resolutions for each band. For \Gaia\ $G$, we consider all systems with separations below 1\arcsec\ unresolved, while we treat wider systems as fully resolved with the secondaries having their own model \Gaia\ sources and include them separately in the model population. For 2MASS $J$, we consider all systems with separations below 2.5\arcsec\ to be unresolved (Kraus et al. 2021, in prep). For systems with separations between 2.5\arcsec\ and 4\arcsec, we treat the system as resolved in 2MASS but do not include the secondary as its own source separately in the model population \citep{2MASS_catalog}. For systems with separations larger than 4\arcsec\ we include the secondary as a separate object in the model population. Synthetic stars that only have one magnitude, exclusively the $G$ magnitude of secondaries in binaries with separations from $1-4$\arcsec, are excluded from the ensemble.

To create a continuous likelihood distribution from this synthetic stellar population, we convolve the $(M_J,M_G)$ synthetic population data with a KDE to calculate a smooth likelihood function in magnitude-magnitude space. We use a Gaussian kernel with a bandwidth of the typical uncertainty in the de-extincted magnitudes for our data, which is roughly 0.2~mag (taking into account magnitude, distance, and extinction uncertainties). We can then evaluate the output of the KDE at a specific combination of absolute $G$ and $J$ magnitude to compute a likelihood. 

Within the MCMC fit itself, we use a linear-flat prior on age between the bounds of our isochrone grid, excluding ages younger than 0.5~Myr and older than 100~Myr. While 0.5~Myr is the lower limit set by the isochrone grid, it also represents the expected duration of the Class 0+1 stages, and hence is the point when stars become easily observable \citep{Evans09}. For MCMC sampled ages within the grid and the ensemble of objects, the value of the posterior distribution is simply the combination of the likelihood for each of the individual objects:

\begin{equation}\label{eq:AgeBayes}
    \log{P(\mathrm{age}|G, K)} = \sum_{i=1}^N \log{\mathcal{L}(G_i,K_i)}
\end{equation}

\noindent where $N$ is the number of stars in the population, $\mathcal{L}$ is the smoothed likelihood function from the KDE, and the log-prior term is excluded as it evaluates to zero. To speed up computation, we pre-compute a grid of KDEs at intervals of 0.01~Myr and use the KDE with an age closest to the sample age within the MCMC in the likelihood calculation. While this discretizes the likelihood space, it is much smaller than the typical uncertainty in the fit age and the difference between KDEs at this age spacing is minimal. This framework is flexible and can be extended to include other parameters for future investigation, such as the IMF itself, binary demographics, or an age spread within a subgroup.

We exclude from all fits objects that: 1) have $A_V>5$, 2) have been previously identified as Class 0 or I, and 3) have spectral types earlier than F5. These objects have less reliable absolute photometry, and are often outliers that negatively affect the quality of the fit.

\begin{figure}
\begin{center}
\includegraphics[width=\columnwidth]{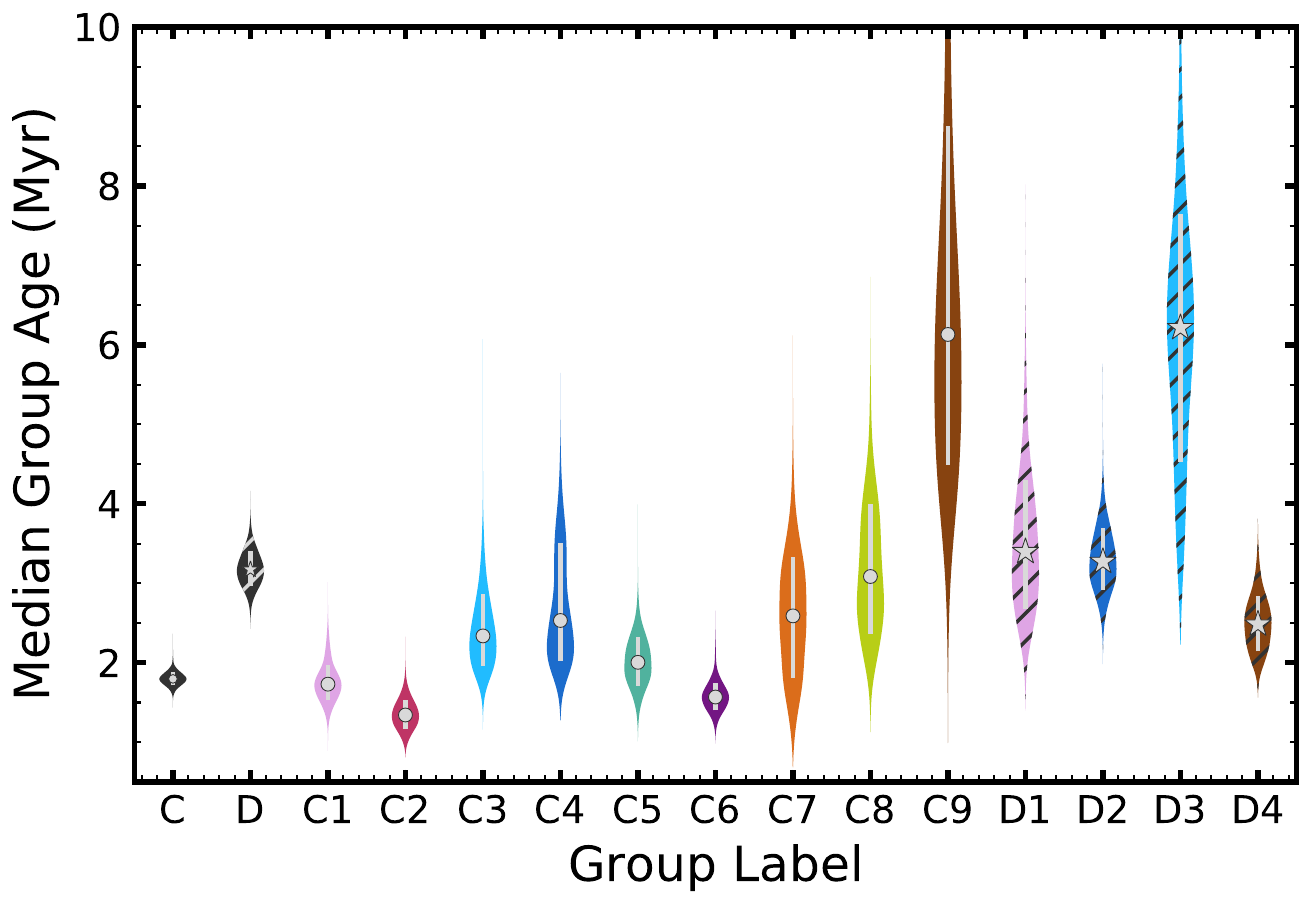}
 \caption{Posterior age distributions of the Taurus subgroups. We exclude Groups \cTen, \dFive, \dSix, and \dSeven, as they only have 3 or fewer objects. The first two posteriors are for ``super-group" fits: fits on all of the core group members together (label C) and all of the distributed group members together (label D). Colors for the numbered groups retain their meaning from Figure~\ref{fig:xyz_min_gmm}; normal fill denotes core groups and hatched fill denotes distributed groups. The markers denote the median value, and the error bars mark the 16th to 84th percentile range. Generally, the distributed groups are older than the core groups, which makes sense as they are removed from the clouds and regions of ongoing star formation. This is highlighted by the roughly 1.5~Myr difference between the Core and Distributed ``super-group" fit results, which is well outside of the corresponding uncertainties. Group \cNine\ is an exception, which makes sense as it is spatially distinct from the core Taurus cloud region, but very tightly-defined unlike the distributed groups. The broadened and slightly multi-modal distributions for Groups \cSeven, \cEight, and \cFour\ are likely due to smaller number of members included in the fit. \label{fig:age_violin}}
\end{center}
\end{figure}

\subsection{The ages of Taurus and its subgroups}\label{subsec:agediscussion}

We have applied the Bayesian formalism of Section~\ref{subsec:agecode} to each of the groups identified in Section~\ref{subsec:gmm_results}, determining their best-fit ages and corresponding uncertainties. We have also fit all core groups together and all distributed groups together to assess the ages of the region at large.  

Figure~\ref{fig:age_code_example} shows the fit output for one of our identified groups: Group \cFive, which is a core group of 29 objects near the central areas of ongoing star formation. The data are shown in both magnitude-magnitude and color-magnitude space along with the KDE of the best-fit median age of $2.01^{+0.32}_{-0.29}$~Myr. The posterior distribution on age, and spatial distribution of the group are also shown. The KDE encompasses the data points well, showing that it accurately accounts for the measurement error and captures the spread in the data. The good match between the group's magnitude-magnitude spread and the KDE broadened solely by observational errors implies that there isn't a need for an age spread. This is not true of all groups, however, and will be discussed later. We also find the upper uncertainty to be larger than the lower uncertainty for all fits, although it is more pronounced at older ages. This correctly reflects that the magnitude-age gradient becomes shallower towards older ages, eventually resulting in a tail towards older ages in the posterior beyond a few Myr. While there are substantial uncertainties in the absolute ages, even as much as factor of 2
\citep[e.g.][]{Naylor09,Pecaut12,Malo14,Feiden16,Rizzuto20}, relative ages are measurable to fairly high precision barring any serious systematic issues shared by many members of a group. A precision relative age measurement is particularly possible with sufficiently large group membership, which is achieved even in the lesser populated groups such as Group \cFive.

Figure~\ref{fig:age_violin} shows the posterior age distributions from the Bayesian formalism for all groups. The adopted ages and uncertainties for each group are presented in Table~\ref{tab:groupprops}. In general, the distributed groups are older than the core groups, except for Group \cNine; this is unsurprising as Group \cNine\ is tightly-defined like a core group but separated from the cloud complex. Groups that have broad, vaguely multi-modal posteriors, such as Groups \cFour, \cSeven, and \cEight, suffer from having fewer objects than other groups. We do not show the outlier groups (Groups \cTen, \dFive, \dSix, \dSeven), as they all have 3 or fewer members which results in poorly constrained ages. There is only modest inter-group age spreads for each type of group (core and distributed), except for Group \dThree\ which is significantly older than the other distributed groups, like Group \cNine\ is to the other core groups. 

Groups \cThree\ and \cFour, which are spatially coincident and appear to be a central core and dispersed halo around it respectively, have ages that are in good agreement. We ran a fit combining the members of these two groups, which forms a well-defined sequence in the CMD. This results in an age that falls between the reported ages for the two groups, with a well-constrained posterior distribution. Along with their kinematic similarities, we conclude that these two groups are linked to the same local star forming event. Future surveys of these combined subgroups, with a more complete sub-census, could investigate population properties, such as the mass distribution, to search for radial dependencies in low-mass star forming events.

The age uncertainty is strongly dependent on the number of objects included in the fit. 
However, we find that the distributed groups have slightly larger uncertainties than the core groups even when accounting for the number of objects included in a fit. We conclude that the increased uncertainty is a result of the distributed groups having age spreads within them. This is consistent with the idea that the distributed groups are conglomerations of two populations of stars: one from previous epochs of star formation that have dispersed, and one from recent and isolated star formation throughout the region.

\subsection{Ages of the core and distributed populations}\label{subsec:supergroup_ages}

To further quantify the age difference between the core and distributed populations, we performed ``super group" fits, which use the Bayesian formalism separately on all members of core groups together and all members of distributed groups together. The posterior distributions of those fits are shown as the first two distributions in Figure~\ref{fig:age_violin}. We find a core group population age of $1.80\pm0.08$~Myr and a distributed group population age of $3.17^{+0.23}_{-0.21}$~Myr. We emphasize that these ages are systematically uncertain in the absolute sense. However, in the relative sense the distributed group population is nearly twice as old as the core group population, with an age 1.5 Myr older, which is well outside of uncertainties. We find that the distributed group population has a larger age uncertainty than the core group population. This is partially due to the fact that there are fewer objects in the distributed than the core ``super group" fits (163 vs. 266). Even after accounting for the number of objects in the fit (assuming uncertainty goes as $\sqrt{N}$), however, the distributed group age uncertainty is twice that of the core group population. We interpret this to mean that the distributed groups have more of an age spread than the core groups. 

\begin{figure*}
    \centering
    \includegraphics[width=\textwidth]{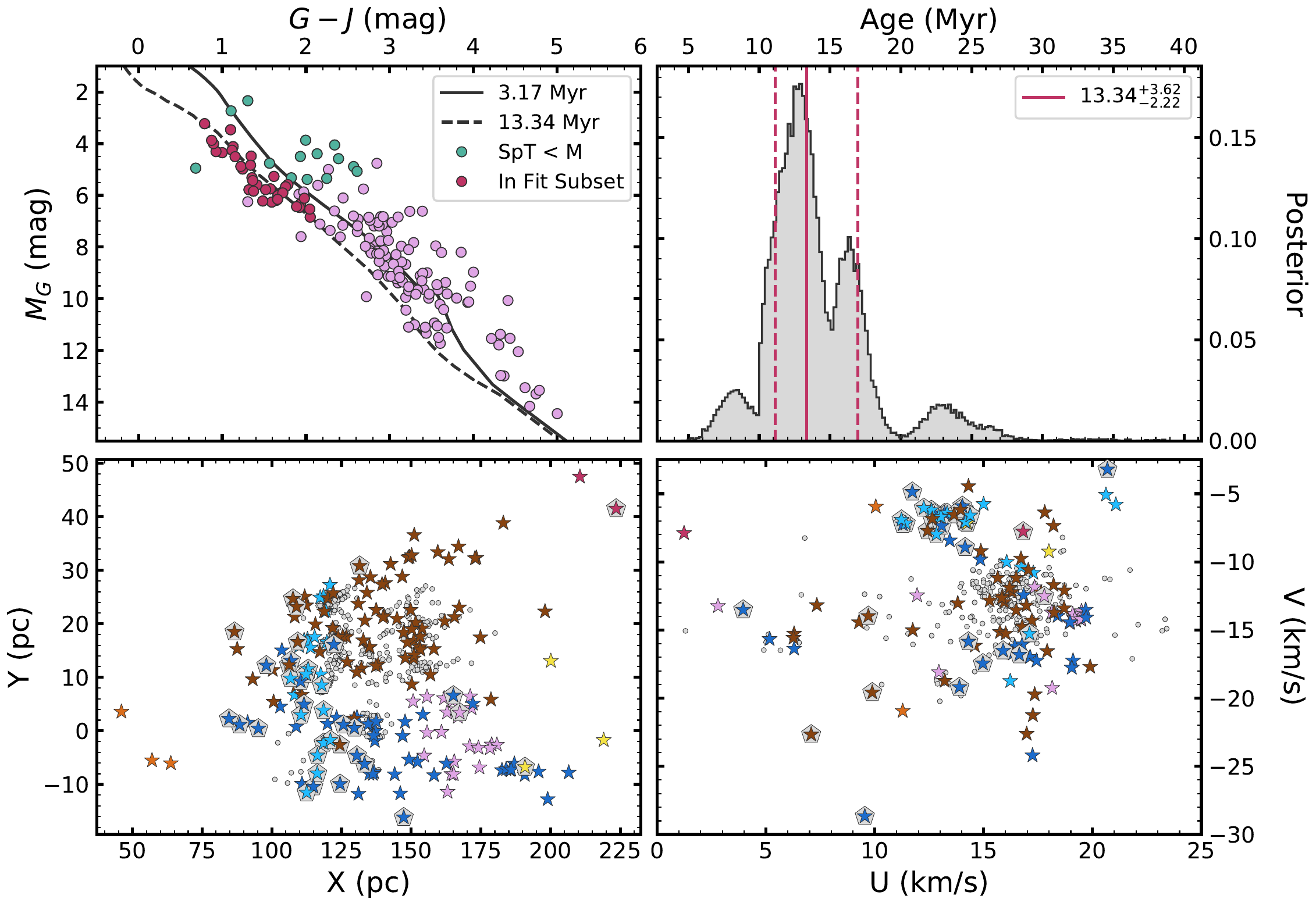}
    \caption{Our investigation into a separate older sub-population amongst the distributed groups. \emph{Upper left panel}: A CMD showing the distributed group members, with separate points plotted for members with spectral types later and earlier than M, and objects older than the overall best-fit age of $3.17$~Myr. The isochrone for the age from the full population fit is plotted as the solid line. The best-fit isochrone for the earlier-type older objects is plotted as the dashed line. \emph{Upper right panel}: The age posterior for the fit of the earlier-type older subset of the distributed groups. The posterior is multimodal, likely due to the outsize influence of individual objects on a fit with a small number of data points. The median best-fit age visually fits the data well in the CMD panel. \emph{Lower left panel}: The spatial distribution of the Taurus region in the $XY$ plane. This plane was chosen as it carries the most differentiation between the older subset of objects and the rest of the distributed population. The older population is marked with background gray pentagons and the core group members are plotted as small gray circles. The older population is closer in $X$ than the rest of Taurus, while spanning the same extent in $Y$. \emph{Bottom right panel}: The kinematics of the older distributed population in the $UV$ plane, which contains most of the differentiation between the populations. The older population is preferentially in a tight location of $UV$ space at slightly lower $U$ and significantly larger $V$ velocities. Older objects that are not members of Group \dThree\ are found throughout $UV$ space. \label{fig:distr_multipop}}
\end{figure*}

The age spread is evident in the distributed population's CMD, which is quite broad, and shown in the upper left panel of Figure~\ref{fig:distr_multipop}. In the CMD, there are two apparent populations with different ages, and these populations are demarcated by spectral type. The M-dwarf population, and a handful of earlier objects, agree well with the fit age of the whole population; this is not surprising as the age fit is driven by objects closer to the peak of the IMF. There is a tight sequence of objects with spectral type earlier than M that is older than the rest of the population. The earlier spectral type preference of this older sub-population is likely due to selection effects: they have been discovered through surveys that preferentially find earlier type objects. To investigate this older sub-population, we performed a separate age fit on objects in the distributed group population that have spectral types earlier than M and are below the $3.17$~Myr isochrone by more than $1\sigma$ in $M_G$. 

The age posterior of this older sub-population is shown in the upper right panel of Figure~\ref{fig:distr_multipop}. We find a best-fit age of $13.34^{+3.62}_{-2.22}$~Myr, with the corresponding isochrone plotted in the CMD in Figure~\ref{fig:distr_multipop}. The posterior is multimodal, which is likely due to higher sensitivity to outlier objects since there are only 36 data points in the fit. Still, the median age visually matches the CMD sequence exactly. The bottom two panels in Figure~\ref{fig:distr_multipop} show the spatial and kinematic distributions of this older population. These objects are preferentially closer than the rest of the Taurus groups, which matches the finding from \citet{Kraus17} that the distributed population is a collection of older, foreground objects. These objects have a wide spread in kinematics, but have an overdensity at $V$ velocities larger than the rest of the Taurus region. 

The spatial and kinematic properties of this older population closely match those of Group \dThree, which largely comprises these older objects. Group \dTwo\ also contributes multiple members to this older population, but it is a relatively insignificant fraction of its full membership, and half of its older objects are not located near the kinematic locus of Group \dThree. Additionally, the spatial distribution and bulk motion of Group \dTwo\ is relatively unchanged if you exclude these older objects. We conclude that Group \dTwo\ is a robust and large group of objects of varying age distributed throughout the Taurus region. The other distributed groups contribute negligibly to this older population. 

The core groups are newly formed, relatively tight clusters of stars that are still near their places of birth. The distributed groups are collections of objects of varying ages, from newly formed to over $10$~Myr, that are related to and distributed throughout the Taurus region. They are a combination of older objects that formed in clusters that have since dispersed, younger objects that have been ejected from their birth site, and in the largest fraction, objects of mixed age that simply formed in isolation from the areas of intense and clustered star formation. The ensemble of core groups could evolve to look like the distributed groups: a conglomeration of stars with similar spatial and kinematic properties that formed concurrently but have since dispersed and erased the small-scale substructure with which they were born.

\begin{figure*}
    \centering
    \includegraphics[width=\textwidth]{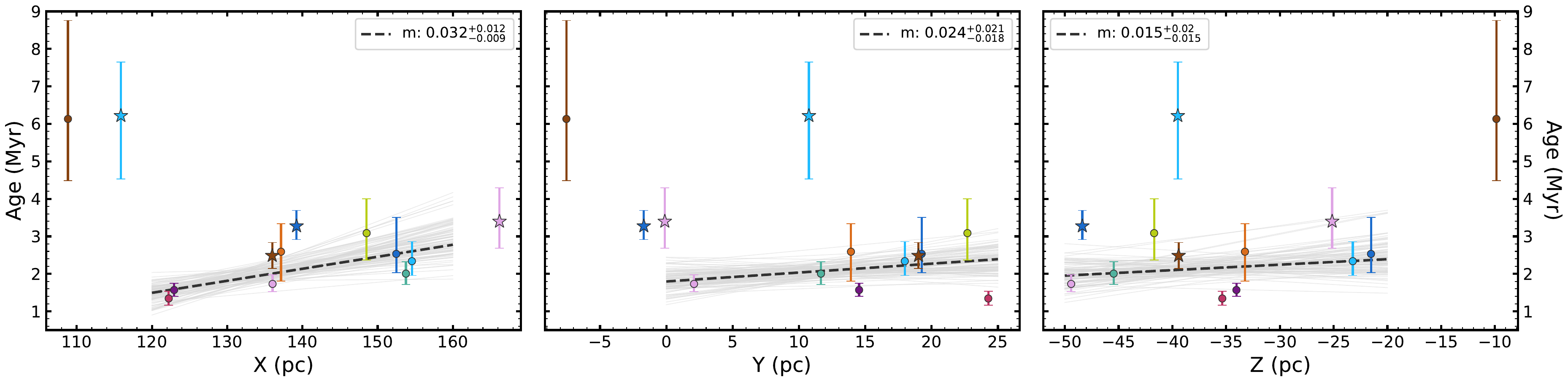}
    \caption{Correlations between the age and galactic location of the Taurus groups, in $X$, $Y$, and $Z$ from left to right. Bootstrap fits of the core groups (excluding Group \cNine) are shown as the dashed lines, with random pulls from the posterior plotted as the background gray lines. The $X$ trend is significant at above the $3\sigma$ level, while the other trends are only just above the $1\sigma$ level. The significant $X$ trend shows that the canonical Taurus region is older with increasing distance, with a gradient of 1~Myr over the full extent of the region. While excluded from the fit, distributed Groups \dOne, \dTwo, and \dFour\ follow the trend. \label{fig:age_xyz}}
\end{figure*}

\subsection{There is an age gradient across the greater Taurus-Auriga complex}\label{subsec:age_gradient}

Lastly, we investigate any trends between the ages of the Taurus subgroups and their galactic positions and velocities. To do this, we fit linear trends to the core groups' ages and bulk position or velocity points. Figure~\ref{fig:age_xyz} shows the median group ages as a function of their mean $XYZ$ galactic positions. We do not include the outlier groups \cTen, \dFive, \dSix, \dSeven\ in the plot. There are no significant non-zero trends in $Y$ or $Z$. The $X$ trend, however, is highly significant at $3\sigma$ above no slope. This means that in the canonical Taurus region, age increases with distance. This is not consistent with the idea that the star formation in the canonical Taurus region was triggered from the direction of the foreground older population, but actually from the other direction. While they are not included in the fit, the other distributed groups (\dOne, \dTwo, and \dFour) do generally follow this trend. We find no significant age trends with $UVW$ velocities.

As an outlier in age, location, and kinematics, observed again in this work, the older foreground population has not been considered to have relation to the main Taurus region. However, we conclude that none of those things preclude a connection between the older population and the ongoing star formation in the canonical Taurus clouds. While it is kinematically distinct from the core Taurus region, it is only slightly more so than would be expected from Larson's Law, within velocity errors. It is also entirely plausible that the population was birthed from a distinct molecular cloud with its own bulk motion, but was a part of a larger network of gas in the galactic neighborhood from which the canonical Taurus clouds formed. The cloud from which this population formed could have became dense enough to trigger star formation before the canonical Taurus region, and as a likely similar low-mass star forming event it never produced a supernova to trigger star formation in the surrounding areas. The members of Group \dThree\ that are in the older foreground are slightly closer in $X$ than the rest of Group \dThree, showing that there is at least some gradient towards younger ages between the older foreground population and the canonical Taurus region. 

Instead, the canonical Taurus region became dense enough to start forming stars from a trigger further away in $X$, independent of the older foreground. If it were a supernova, it could be the source of the energy injection into the core group population at a scale of $\sim15$~pc seen in the inter-core-group VSF in Figure~\ref{fig:vsf}.

\section{Discussion and Conclusions}\label{sec:discussion}

The star formation history of the greater Taurus region is complicated. We conclude that there is a distributed population of slightly older objects in the Taurus region, coincident in both space and in kinematics. This population might represent an earlier generation of star formation that resembled the current generation and has since dispersed, or it might be a collection of objects formed in a sparse, distributed configuration before the clustered groups of young stars near the clouds of ongoing star formation. Regardless, this means that the Taurus cloud has been actively star forming for at a minimum of 5~Myr, which is still within the expected lifetime of a molecular cloud. Indeed, the presence of disk-hosting mid-M stars without recognizable lithium, such as StH$\alpha$ 34 \citep{White05,Hartmann05}, argues the total duration might exceed 10 Myr. We also recover the significantly older population identified in \citet{Kraus17}, with an age of roughly 15~Myr, but it is spatially and kinematically distinct from the core Taurus region. However, this does not preclude a relation between the two, as the velocity difference is within expectation from Larson's Law and it seems likely that Taurus is a part of a larger, extended star formation event in the galactic neighborhood.

The ages of the core groups are roughly equal to, or at least not significantly younger, than their relaxation timescales. We conclude that the core groups have had time to at least modestly dynamically relax, and that the slope of the intra-core-group VSF is shallower than that of the parent molecular cloud's turbulent power spectrum due to this dynamical evolution. We interpret the larger slope of 0.35 for the inter-core-group VSF at the largest scales to be a true approximation of the turbulent power spectrum of the parent molecular cloud. This would indicate a Larson-like subsonic turbulence. As the distributed groups are only modestly older than the core groups on average, we interpret the difference in the slope of their corresponding VSFs to be a result of their modes of star formation. We conclude that the slope of the intra-distributed-group VSF is an approximate lower limit on the initial power spectrum of the parent molecular cloud. The distributed groups are a mix of objects that have formed in isolation, whose velocity spectrum should reflect the initial turbulent structure, and older stars that formed in groups, which if analogous to the core groups would have had time to dynamically relax before dispersing, contributing a shallower VSF slope. This is consistent with the intra-distributed-group VSF slope being moderately shallower than the inter-core-group VSF slope. The higher velocity dispersion in the distributed groups across all scales is likely due to dispersion increasing with age, and the more disparate origins of the objects in the distributed groups. 

The exquisite detail with which we can now map the structure of star forming regions like Taurus is key to our understanding of the evolution of stellar associations and the interpretation of the phase space distribution of older stellar populations. The many nearby young moving groups, and more that will be discovered with the ever-better \Gaia\ catalogues \citep{Torres08,Mamajek16,Gagne18b}, are thought to have been formed in more distributed groups such as Taurus \citep{Kraus08}. If Taurus-like regions are common, then from this work we expect young moving groups to have formed in both clustered and distributed modes. For example, the TW Hydrae Association may have formed in one or more groups along a filament without much distributed star formation, like a subset of the Taurus core groups, thus producing its elongated spatial structure \citep{Weinberger13}. In contrast, the Tuc-Hor Association may have formed from nearby filaments aligned in $Z$ with distributed star formation throughout, as though the Taurus region were rotated in $XYZ$, to produce its sheet-like structure \citep{Kraus14}. Knowledge of the distribution of stars at formation is needed to determine ages through kinematics, such as through the forward modeling technique in \texttt{Chronostar} \citep{Chronostar}.

To summarize our findings:

\begin{itemize}
    \item We have identified 17 spatially-distinct groups in the greater Taurus-Auriga star forming region by fitting a Gaussian mixture model to the most expansive and inclusive census of the region to date. These groups feature a dichotomy in number density: with the higher density ``core" groups residing near the molecular clouds and the lower density ``distributed" groups spread throughout Taurus.
    
    \item Care must be taken when using \Gaia\ astrometry to investigate the spatial distribution of stellar populations. For example, the groups in Taurus appear elongated along the line-of-sight. We fit the groups' spatial distributions to account for individual object covariance matrices and include an error inflation term as \Gaia\ errors are under-reported. We find that the line-of-sight elongation is an artifact of \Gaia\ astrometric error being almost exclusively in the distance to objects, and that nearly all groups are equal in depth and spread on the sky. We also find that \Gaia\ astrometric errors are under-reported by roughly 12\%.
    
    \item The groups in Taurus have largely similar bulk kinematics, with differences being more pronounced for groups removed from the clouds. There are two coherent sub-populations amongst the core groups that track position beneath the galactic plane. The distributed groups have larger velocity dispersions, which follows expectations from Larson's Law. The two main distributed groups that cover most of the Taurus region have bulk motions in agreement with the core Taurus region, meaning they are unquestionably associated. 
    
    \item The velocity structure of the stellar population is similar to Larson's Law, following a power law with increased dispersion at wider separations. We find the signal of an underlying turbulent spectrum with a power law index in agreement with the original formulation of Larson's Law \citep{Larson81}, which is lower than currently accepted values. The core groups are small enough to have begun dynamically relaxing, leading to a shallower slope for the velocity structure within core groups.
    
    \item We derived isochronal ages for the 17 groups in Taurus, and find a spread from 1.3 to 6.2~Myr in the median age. The distributed groups are older than the core groups, which is consistent with the core population's location near the clouds of ongoing star formation. There is a trend in the core population for group age to increase with $X$ coordinate, which we conclude to mean that the current epoch of star formation in the Taurus region was triggered from beyond it in the galaxy. 
    
    \item There is a sub-population in the distributed groups (largely in Group \dThree) that is significantly older than the rest of the region with an age of roughly 13 Myr. It is closer than the rest of the Taurus region and has a distinct bulk velocity. While it is clearly distinct from the core Taurus region, we cannot say with certainty that this population is unrelated: its velocity separation is not unexpectedly high given its distance from Taurus, and it could be the result of star formation in a separate molecular cloud in the galactic neighborhood of Taurus. However, it is still a small fraction of the distributed Taurus population, and there does exist a slightly older distributed population that shares the spatial and kinematic properties of Taurus.

\end{itemize}

Our picture of the entire Taurus region is still incomplete, and our understanding of its star formation history remains an unfinished tapestry. The census presented here represents a significant fraction of the Taurus region, but further studies will be necessary to find more members and obtain 3D kinematics to enable full 6D clustering. In particular, conclusions about the older and distributed populations are biased by incompleteness, and a more thorough search for members of those groups is needed. Recent work performing all-sky searches for young stars will be crucial in this effort, and to identify more related young populations in the Taurus galactic neighborhood \citep{Zari18,Kounkel19b,McBride20,Kerr21}. Moving forward in the \Gaia\ era, we will be able to place the story of specific regions like Taurus into the broader context of star formation in our galaxy at large.

\acknowledgements
We thank the anonymous referee for insightful comments, and for help in improving the clarity of the manuscript. DMK was supported in part by NASA grant 80NSSC18K0405 through the ADAP program and by NASA grants 80NSSC18K0390 and 80NSSC18K0392 through the K2 mission. ACR was supported as a 51 Pegasi b Fellow though the Heising-Simons Foundation.

This work has made use of data from the European Space Agency (ESA) mission \Gaia\ (\url{https://www.cosmos.esa.int/gaia}), processed by the \Gaia\ Data Processing and Analysis Consortium (DPAC, \url{https://www.cosmos.esa.int/web/gaia/dpac/consortium}). Funding for the DPAC has been provided by national institutions, in particular the institutions participating in the \Gaia\ Multilateral Agreement.

This publication makes use of data products from the Two Micron All Sky Survey, which is a joint project of the University of Massachusetts and the Infrared Processing and Analysis Center/California Institute of Technology, funded by the National Aeronautics and Space Administration and the National Science Foundation.

The authors acknowledge the Texas Advanced Computing Center (TACC) at The University of Texas at Austin for providing high-performance computing resources that have contributed to the research results reported within this paper.

\software{Astropy \citep{Astropy13, Astropy18}, astroquery \citep{astroquery}, emcee \citep{emcee}, ipython \citep{ipython}, jupyter \citep{jupyter}, matplotlib \citep{matplotlib}, NumPy \citep{numpy}, pandas \citep{pandas}, saphires \citep{saphires}, scikit-learn \citep{sklearn}, SciPy \citep{scipy}}

\clearpage

\bibliography{taurus}

\end{document}